# A Survey on Machine Learning for Optical Communication [Machine Learning View]

M. A. Amirabadi

Email: m_amirabadi@elec.iust.ac.ir

**Abstract-** Machine Learning (ML) for Optical Communication (OC) is certainly a hot topic emerged recently and will continue to raise interest at least for the next few years. The rate of research development in this area is growing very rapidly. Novelty of this research direction resides mainly in the peculiarity of the application field, rather than in the methodological approaches, which are (at least up to now) state-of-the-art ML algorithms. Reviewing the literature shows that many of the ML algorithms have not yet been used in this area, and many of the OC applications are not considered yet, which reflects the fact that the research topic is pristine. Accordingly, tutorial investigations are quiet necessary in this filed to help researchers be aware about the last progressions and cavities of this field. Although several tutorials have been released recently, they considered this topic from OC view, and neglected ML view. However, it is required to have an investigations about the ML algorithms used in this subject. Accordingly, for the first time, this paper reviews ML for OC literature from ML viewpoint. This view could be really helpful because only OC experts work on ML for OC, and they are not ML experts, so it could really help them to have a comprehensive view on the ML subjects implantable in OC. It has worth to mention that compared with other works, this survey reviews much more investigations; therefore, it has more generality, and gives the reader to have a comprehensive overview on this topic.

I. Introduction

During the time, humanity demands grew very rapidly in communication system applications. Hence, many research fields appeared and continuously focused on subjects such as increasing bandwidth, rate, security, reliability, etc. In spite of vast investigations and solutions, traditional communication systems could not satisfy high humanity demands, especially in critical points such as bottlenecks. Meanwhile, a new field in science appeared under the name of Photonics, which was the interaction of photon and material, and fundamentally had some features that could deserve the mentioned demands. The development of interdisciplinary researches yielded to the emergence of Optical Communication (OC) systems. These systems were successful somehow, but some of the former problems still remained; e.g. these systems did not have sufficient reliability. On the other side, in Computer Science, a new field was growing exponentially that made devices to be more flexible, and therefore reliable; it was named Machine Learning (ML), which in recent years transformed to one of the most convenient ways for caring about communication systems and networks. In these systems, an algorithm (machine) attempted to predict and eliminate defects by learning the properties of the system. For example, ordinary detectors were replaced by ML, because ML improved the accuracy and performance of the system by creating a nonlinear detection boundary. In addition, ML computation was less than ordinary detectors, and this was the way that the ML was officially introduced into OC systems.

The main idea of complementing ML with OC is the computational complexity of analytical/numerical calculations of OC systems/networks. The Machine can be likened to a body that takes its breath away from its entrance. In fact, the machine takes in input color through learning, and so it can be used to predict an input that has not yet been given. The learning of the machine, itself has progressed a lot, but its applications in OC are still at the beginning. Wavelength Division Multiplexing (WDM), Orthogonal Frequency Division Multiplexing (OFDM), and Spatial Division Multiplexing (SDM) as well as different modulations including Differential Phase Shift Keying (DPSK), Amplitude Phase Shift Keying (APSK), Pulse Amplitude Modulation (PAM), On-Off Keying (OOK), and Quadratic Amplitude Modulation (QAM) are used in ML for OC literature. In addition, many OC system/network subjects such as performance monitoring, phase modulation, fault detection, predictive maintenance, synchronization, fiber nonlinearity equalization, wavelength assignment, blocking probability calculation, network performance evaluation, dynamic bandwidth allocation, bit error rate prediction, quality of transmission calculation, network optimization, lightpath request, constellation shaping, network monitoring, virtual network slicing, adaptive equalizer, knowledge-defined networking, are investigated in ML for OC literature. Furthermore, various ML algorithms such as (in supervised category [1-112]:) Support Vector Machine (SVM) [1-31], Artificial Neural

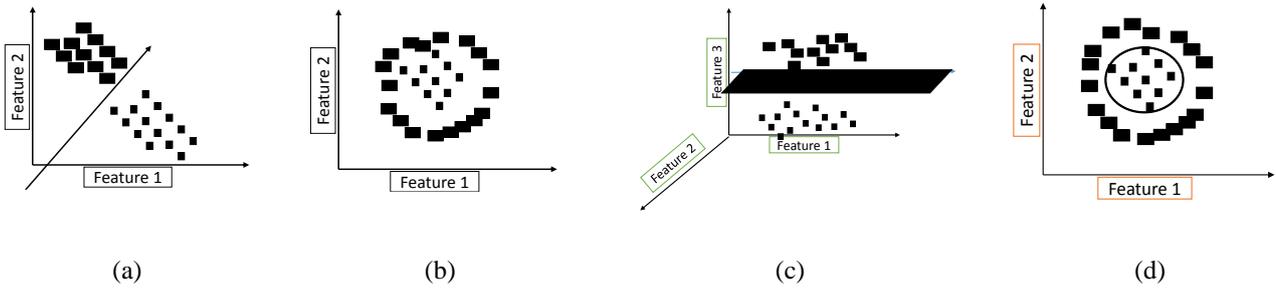

*Fig.1.a. SVM, b. nonlinear data structure, c. dimension insertion, d. Kernelized SVM.*

Network (ANN) [32-72], Deep (DNN) [73-89], and Convolutional Neural Network (CNN) [90-98], k-Nearest Neighbors (KNN) [99-102], Random Forest [103-105], Regression [106-112], (in unsupervised category [113-132]:) k-means [117-126], Principal Component Analysis (PCA) [127-132], (in Reinforcement Learning category [133-151]:), Q-Learning [146-149], and Deep Reinforcement Learning [150-151]are used in the referenced papers of this manuscript.

### A. Related works

Recently, several tutorials released in ML for OC: [152], with a network layer view, provided a comprehensive survey on investigations in AI in OC networks. It first reviewed applications related to physical layer, including optical performance monitoring, nonlinearity mitigation, and estimating quality of transmission (QoT). Then reviewed applications such as optical network planning and operation in both transport and access networks. [153], again with a network layer view, reviewed the application of ML for OC and networking. [154], again took a network view, and reviewed works on ML for OC network and discussed algorithm choices, as well as data and model management strategies. [155], again with a physical layer view reviewed ML investigations on CO-OFDM and focused on practical aspects and comparison with the DSP techniques. Also a comparison is done over computational complexity of ML and DSP. [156] described the mathematical formulations of ML algorithms from communication theory view; further it reviewed some of investigations on ML for OC with a physical layer view.

Although there are published several surveys on ML for OC, however, none of them had ML view, and number of reviewed works in them were limited. Considering that ML for OC investigations require knowledge and awareness in both ML and OC aspects, and the fact that there is lack of investigation that clarify the ML techniques used in ML for OC, this paper aims to fill this gap, and proposed a comprehensive investigation over a lot of released works in this field. The rest of this paper is organized as follows: sections II, III, and IV are devoted to supervised learning, unsupervised learning, and reinforcement learning. There is a lack of semi-supervised learning research in this area. Section V is the conclusion of this work.

## II. Supervised Learning

Supervised learning uses a pair of data sets (e.g. $x, y; x \in X, y \in Y$), which one of them ($y$, or label) is the mapping of the other ($x$, or feature), and the aim is to find the mapping between them (in training phase). Accordingly, a cost function (e.g. MMSE or cross entropy) should be defined based on the pair mismatches, and minimizing (by e.g. gradient descent algorithms). The applications of supervised learning include pattern recognition (classification) and regression (function approximation). Actually, the learning in supervised form is done with a teacher that provides continuous feedbacks and guidance on the accuracy of solutions obtained.

### A. Support Vector Machine

**Definition:** A Support Vector Machine (SVM) is a discriminative classifier, which based on the input labeled data generates an optimal separating hyperplane that divides input data in two classes (Fig.1.a). There is no separating hyperplane for nonlinear combination of input data classes (Fig.1.b), so, the data should be mapped to higher dimensions to find the separating hyperplane (Fig.1.c). This technique is called kernel that acts as a nonlinear SVM (Fig.1.d).

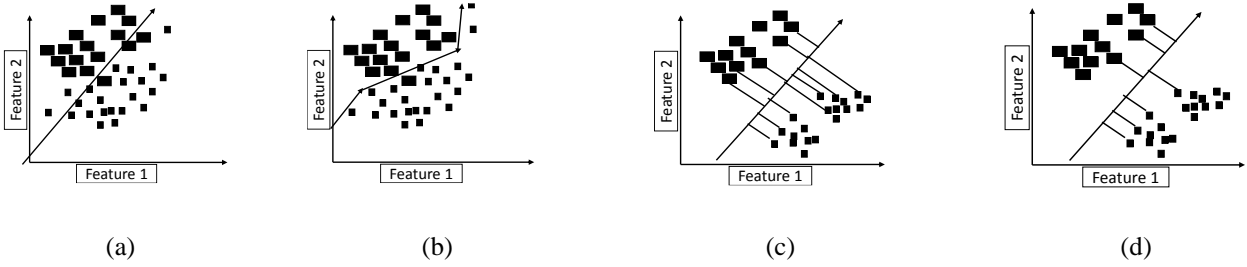

Fig.2. a. overlapped data sets, b. higher SVM model, c. high regularization, d. low regularization.

Depending on the input data type, there are various kernels that could be used (e.g., linear, polynomial, exponential, etc.). Linear kernel calculate separating hyperplane in lower dimensions, while polynomial and exponential kernels calculate separating hyperplane. These three kernels are in the following form:

$$K(x, x_i) = a_0 + \sum_i a_i <x, x_i>, \qquad (1)$$

$$K(x, x_i) = a_0 + \sum_i a_i (<x, x_i>)^d, \qquad (2)$$

$$K(x, x_i) = exp(-\sum_i a_i(x - x_i)). \qquad (3)$$

where $a_0$ and $a_i$ are defined based on training data, $<.>$ is dot production, $x$ is new input, and $x_i$ is support vector. Increasing dimension is only useful for nonlinear data structures; but for situations that two data sets have overlap, the degree of SVM model should be increased. So, two options are available; tolerating some outlier points (Fig. 2.a) or perfect partitioning (Fig. 2.b). The second one is called regularization trick, which uses a regularization parameter to tell the amount of avoided misclassification of each training example. Large regularization parameter (Fig. 2.c) has a smaller-margin hyperplane if that hyperplane does a better job of getting all the training points classified correctly, and an inverse will be held for small regularization parameters (Fig. 2.d). Selecting a good margin (separating line), which could separate classes better, is very important. A good margin allows the points to be in their respective classes without crossing to other class (Fig.3.a), and a bad margin is reverse (Fig.3.b) [1].

**Fiber OC:** The SVM, due to its simplicity, is one of the most widely used algorithms in ML for OC. SVM is shown to be powerful for combatting NLPN [2-4], laser phase noise [5], modulator nonlinearity [6], phase skew and modulator imperfections between in-phase and quadrature [5], fiber Kerr effect, Amplified Spontaneous Emission (ASE) noise, and dealing with linear/non-linear channel/signal detection [7]. SVM could be used anywhere that exist two or multiple distinct groups of data. For example for classification of the received bit sequences to 0 and 1, or classification of normal fiber and fault encountered fiber [8]. Its procedure is simple, it collects data from the distinct categories and learns how to classify them from extracted features. Generally speaking, SVM has lower complexity [9], because it does not consider the hole data, it just searches for hyperplane based on marginal data. Furthermore, with SVM, the redundant features could be deleted, so, the computing cost and complexity would be reduced. Although, SVM leads to the transmission delay due to use of the training data, it has low delay in the testing phase (when separating hyperplane is defined). In addition to complexity reduction, SVM could do well some applications, in which traditional techniques are not effective, e.g. forming an effective fault diagnosis, but, SVM solves it properly by transforming the classification problem into a nonlinear programming problem [8].

Fortunately, despite other ML algorithms, there are many SVM based investigations in ML for OC, so these works could be classified in a better way. One of the most widely used applications of SVM in OC is detection, which has mostly shielded on Fiber OC, and there are few implementations in WOC applications. Actually, SVM is a binary classifier, which can only be used for binary signaling not M-ary signaling. In order to solve this issue, should construct $M$ hyperplane (instead of one hyperplane), one of the efficient M-ary SVM solutions is the bit-based SVM, which only needs $\log_2(M)$ SVMs (hyperplane) for M-ary signaling detection [10]. Another M-ary SVM technique is decision tree, which transforms an M-ary and complex multi-class classification problem into binary and multi-layer classification problem (Fig.4.a, b) [11]. SVM does not require information about the transmission link and has a relatively large improvement especially in high nonlinearity [3], and it is shown that M-ary SVM can perform better than ML detection with back rotation [4].

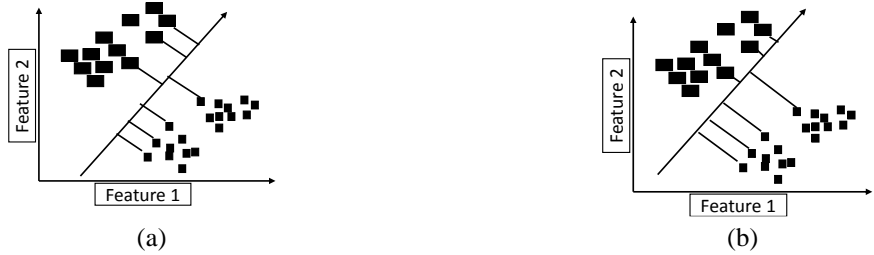

*Fig.3. a. good margin, b. bad margin.*

The higher order modulations (dense constellations) have high capacity, and spectral efficiency, but are sensitive to various faults, and cannot be classified by linear decision boundary (binary classification). Noisy signals have always worse constellations, because they are widely scattered and their boundaries are ambiguous. SVM is able to directly de-map the rotated constellation and has low complexity for making decision. For noisy constellations which are highly overlapped, the standard SVM could hardly create proper decision boundaries, the available extended SVMs are somehow complex and require simplification. Carrier less amplitude phase (CAP) modulation is good for short-range communication links. Results indicate that SVM based detector [12], and equalizer (based on constant module algorithm (CMA)) [6, 13] in high-density CAP OC, has performance improvement over traditional techniques [12].

In addition to detection, SVM is widely used for equalizing linear and nonlinear effects of fiber propagation media [14]. Convolutional equalizers could not compensate these issues because of relative intensity noise and mode partition noise of vertical cavity surface emitting laser (VCSEL) based optical interconnect links. This caused especially that higher order modulations be practically limited. However, this problem solved by M-ary SVM methods such as one versus rest (Fig. 5.a), symbol encoding, binary encoding (Fig. 5.b), and in-phase and quadrature components (Fig.5.c) [15]. The SVM based nonlinear equalizers (NLE)s are only implemented in Fiber OC [14]; however, they could be considered for the same task in WOC systems in future researches. The blind spot of SVM is the insufficiency of features; this issue could be solved by feature enhancement methods such as sync interpolation. However, feature enhancement decreases SVM speed; this issue could be solved using Newton based SVM, which reduces classifier complexity, and computational load [16]. An SVM-based NLE could be applied to single/multi-channel WDM and OFDM [17] systems for increasing nonlinearity tolerance in long-range communication [18]. It could tackle deterministic fiber nonlinearity and the interaction between nonlinear effects and stochastic noises, as well as inter-subcarrier nonlinear crosstalk effects. The fiber effects are really hard to be modeled, therefore, the detection and equalization should be done blindly or approximately. NLE/Blind NLE reduces the fiber nonlinearity penalty and increases data rate compared with inverse Volterra series transfer function (IVSTF) NLE. Combining the SVM cost function with Sato and Godard error functions and then minimizing the obtained cost function using iterative algorithms, e.g. least squares (LS) [17] or weighted least square (WLS) [19], releases a new version of SVM which could be used for Blind NLE. This technique better reduces nonlinearity compared to IVSTF NLE, and is a bbeter choice for multi-channel configuration.

The investigated SVM based detection, and equalization techniques used binary or M-ary SVMs. SVM is a linear classifier, and is not suitable for nonlinear input features (e.g. in classifying filter cascading effects); for solving this issue, a mapping should be done from the feature space to the so called kernel space, in which features could be separated linearly [20]. Kernel based SVM has recently become more popular, because it helps linear algorithm to be deployed non-linearly, and obtain significant performance improvements with higher complexity. It has a flexible decision boundary, because any desired boundary shapes can be created utilizing Gaussian radial basis function (RBF) [21]. Besides SVM, other kernel methods include kernel linear discriminant analysis (Fig.6.a.), kernel K-means clustering (Fig.6.b.), kernel PCA (Fig.6.c), and kernel online learning (Fig.6.d.) [22]. However, due to the youth of the subject kernel based SVM is only investigated in combating filter cascading effects. Filter cascading leads to enhance stop band rejection and steepness of transition band. However, they affect edge sub-channels of the OFDM super-channel. It is important to identify and compensate them; previously, these were done by phase-conjugated subcarrier coding based nonlinearity compensation technique, now, these could be done using a complex-valued kernel based SVM [23]. In addition to compensation of filter cascading effects, complex-valued SVM could also be used in order to compensate stochastic induced nonlinear cross-talk effects [24]. The same as SVM, kernel based SVM has better performance than traditional direct decision method.

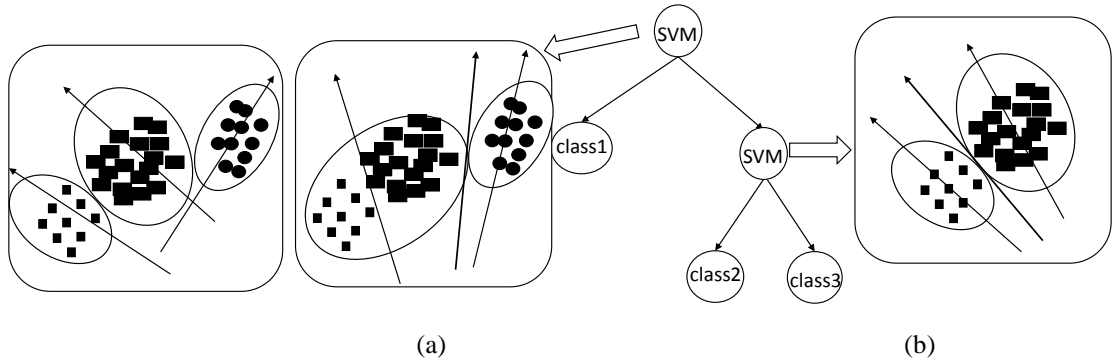

*Fig.4. a. Multi class TWSVM, b. Binary decision tree TWSVM.*

In addition to use of techniques such as bit based or kernel based SVM, which are related to the data structure, some other techniques could be used (for improving performance) that are related to the structure of SVM. Actually SVM has some hyperparameters, which tuning them would result in better performance for the same data. One of the well-known techniques is Genetic algorithm (GA) (Fig.7.a), which despite its performance, has high computational load, and dependent on the situation (e.g. time demanding applications), might not be preferable [25]. Furthermore, for improper fiber nonlinearity, in which the fiber nonlinear interaction and ASE noise expand considerably, the effect of SVM and GA-SVM are similar. Other tuning techniques such as manual search, grid-search, and random search (Fig.7.c) in cross-validation (Fig.7.b) phase, the optimal parameters of SVM could be selected.

The bit, kernel, PCA, tuning based SVM, all have one important barrier, and that is un-flexibility on number of support vectors. Regularized SVM controls number of support vectors and margin errors by adding a regularization parameter to the cost function [26]. The same as SVM, this could be solved using the dual problem; first an equivalent dual formulation for regularized SVM should be constructed, then solved by a robust regularized SVM based on lower-upper decomposition with partial pivoting.

The discussed SVMs all assumed input vectors to be independent, so, they are not appropriate for dealing with inter symbol interference (ISI) applications. For considering this issue, decision feedback based SVM could be used, which considers the temporal correlation between input vectors in the presence of ISI. Further the pattern space viewpoint of the problem illustrates benefit of a bank of SVM. This technique could robustly address nonlinearities in the presence of ISI [27].

**Network:** After detection and equalization, the third important application of SVM is optical link failures prediction/ detection, as well as quality of transmission (QoT) estimation [28]. Because SVM could learn optical link properties with a few training data, and without any prior information or heuristic assumptions. The results indicated efficiency of SVM in reducing the computation time. Detection the intrusion of environmental events while keeping the efficiency is important in perimeter intrusion detection in optical links. In this situation, common extracted features might not be so helpful, because they might be contaminated with noise. Usually, in this situations PCA could be a better choice for extracting features. PCA reorders the features based on their variance, and removes the low variance features assuming them as noise. The PCA based SVM could be applied in perimeter intrusion detection [29]. In addition, an accurate equipment failure risk prediction effectively improves the optical network stability. In addition to traditional methods, such as double exponential smoothing, ML algorithms, such as SVM could be effectively used to improve the accuracy of models for failure prediction, in an optical network [30].

**WOC:** ML techniques are at the beginning of deployment over WOC systems; this is simply because of lack of joint experts on WOC and ML. SVM is mostly shielded on Fiber OC, and there are few implementations in WOC applications such as VLC [9], and FSO [31]. The WOC signaling rates are typically Gb/s, and duration of link atmospheric turbulence is on the order of millisecond. The SVM detection could provide a large data block and proportionally a high bit rate [9, 31].

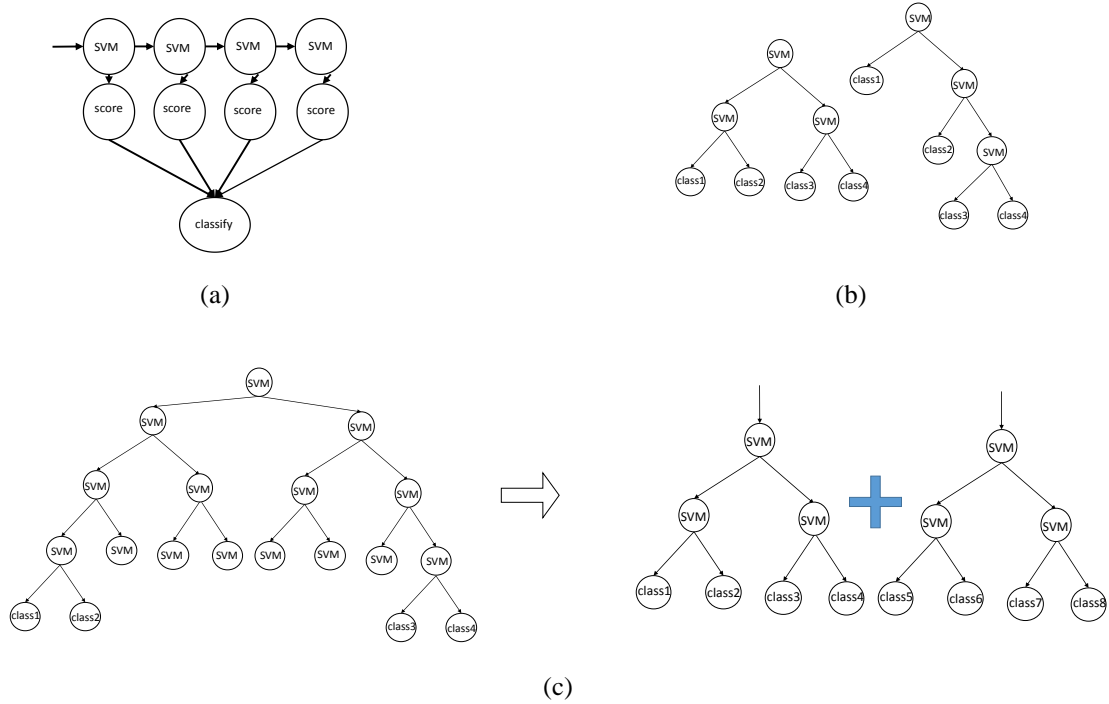

*Fig. 5. SVM based on a. one versus rest, b. binary encoding SVM, c. in-phase and quadrature component;*

### B. Artificial Neuronal Network

**Definition:** An ANN is a member of NN community, which its idea is taken after the working of neurons in the human brain (Fig.8.a). ANN is a variety of ML which comes under the shelf of Artificial Intelligence. ANN, due to its results, is one of the mostly used ML algorithms in OC applications. The basic idea of ANN is taken from brain neural networks, a collection of some neurons (artificial neurons) that are connected to gather. At each connection (neuron) input signals from other neurons are processed and fired by a non-linear function of the sum of input signals. Each connection link has a weight that adjusts while training. The received summed signal at each neuron is compared with a threshold and sent only while aggregating it (Fig.8.b).

ANN is composed of interconnected nonlinear activation units that take multiple inputs and create single output. The input $x = [x_1, x_2, \ldots, x_n]$, weight $w = [w_1, w_2, \ldots w_n]$, bias $\theta$, as well as activation function $f(.)$ are parameters required for calculating the output of a single neuron. If the output is greater than a threshold, then the neuron fires and its output $o$ becomes as follows:

$$o = f(\sum_{i=1}^{n} w_i x_i - \theta) \qquad (4)$$

Simplifying the above function by considering $x_0 = -1$, and $w_0 = \theta$, (4) becomes as:

$$o = f(\sum_{i=0}^{n} w_i x_i) \qquad (5)$$

This procedure is repeated at next (output) layer for each neuron, the aim is adjusting the weight vector and bias such that the transmitted symbol and the output of the ANN be the same. For this purpose a loss function should be defined as:

$$L(w) = \frac{1}{K}\sum_{k=1}^{K}[l^{(k)}(s, \hat{s})], \qquad (6)$$

where $K$ is the Batch size, $s$ and $\hat{s}$ are target and output of the ANN, respectively. Adjusting the weights of an ANN requires solving the optimization problem that is not linear or polynomial; so, it cannot be solved precisely and must be approximated. This could be done by iterative algorithms such as gradient descent methods. However, heuristic algorithms such as GA, due to the complete search result in better solutions [32]. Furthermore, some techniques linearize the constraints and objective functions at a specific point by derivation and partial derivation for some cases [33]. Considering the iterative methods, the process will be stated from an initial point and iteratively declares the updated weight vector. The weight vector updates steepest descent direction of the loss function, which is the derivation direction. In sum, the weight updating formulation becomes as follows:

$$\boldsymbol{w}^{(j+1)} = \boldsymbol{w}^{(j)} - \eta \nabla_{\boldsymbol{w}} \tilde{L}(\boldsymbol{w}^{(j)}) \tag{7}$$

where $\eta$ is the learning rate (step size) [34].

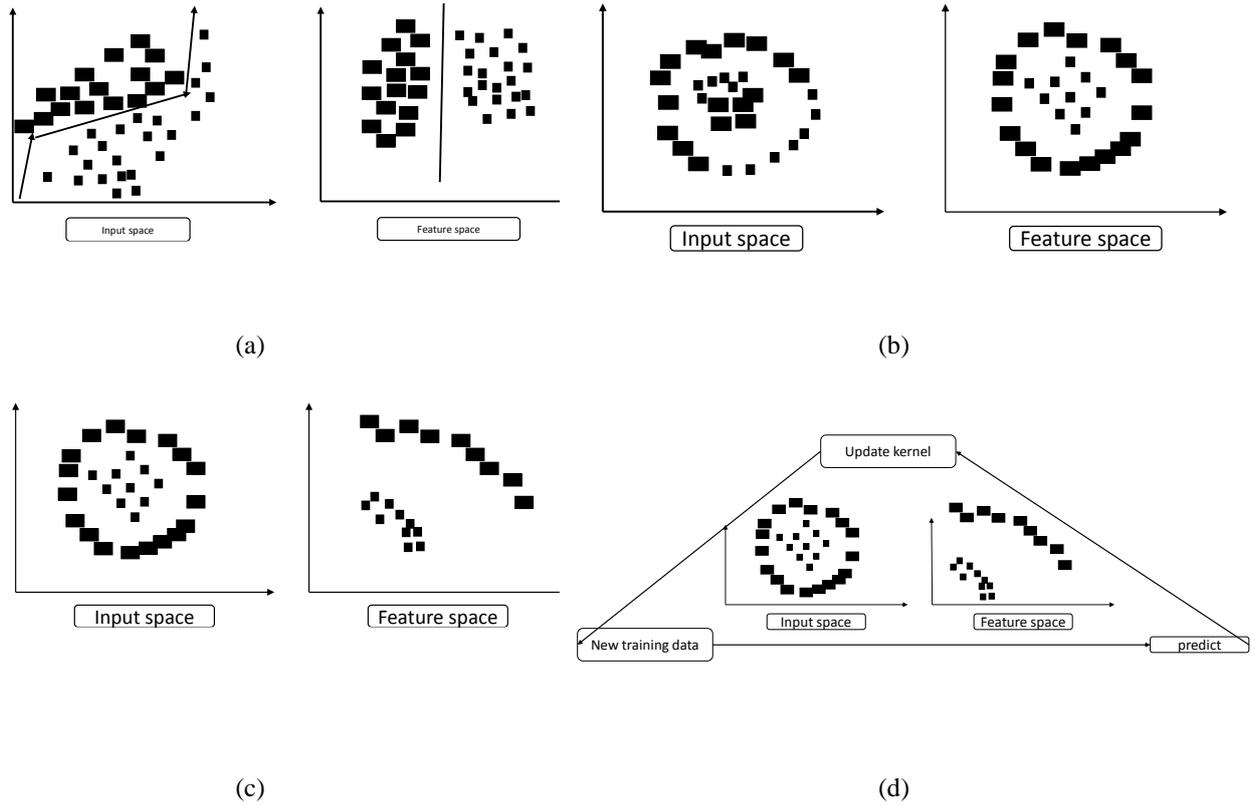

Fig. 6. a. kernel LDA, b. kernel K-means, c. kernel PCA, and d. kernel online learning.

The ANN is a framework of ML that derives complicated relationships between input and target. It learns a specific task by using the label and features without being manually programmed with rules. For example, in bit detection, ANN learns to identify a received bit of data to be 0 or 1 by using example bits that are manually labeled as "0" or "1"; then, the learned ANN could easily identify other bits without prior information by just identifying relationship between input and output. The idea of ANN could be implemented in both electrical and optical domains. In electrical domain, the implementation is the standard ANN implementation. However, ANN could be practically implemented in optical framework (for processing QPSK labels to rout photonic label) [35]. In this framework, neurons are optical amplifier and phase shifter, activation functions are optical nonlinear threshold devices, and connections are optical waveguides. ANN is proven to be very efficient as a classifier. Processing of time-dependent high-speed signals is hard, especially while considering nonlinearly. In these situations, an optical ANN, in which detection is converted to pattern recognition, could help classification of distorted signals [36].

**Fiber OC:** ANN is able to derive complex nonlinear relationships between output and target, therefore, it would be better to be used for complex tasks such as equalization. The same as SVM, most of ANN implementations are related to Fiber OC, ANN NLE could effectively combat fiber effects discussed in the last section [37]. The ANN is a dead body that takes breath by minimizing a loss function (adjusting its weights) using iterative algorithms, or K-means clustering, or least mean square algorithms (generally, iterative algorithms is mostly used for this aim) [38]. The ANN NLE has a very wide implementation scenarios in ML for OC investigations, in the following, these structures are briefly reviewed.

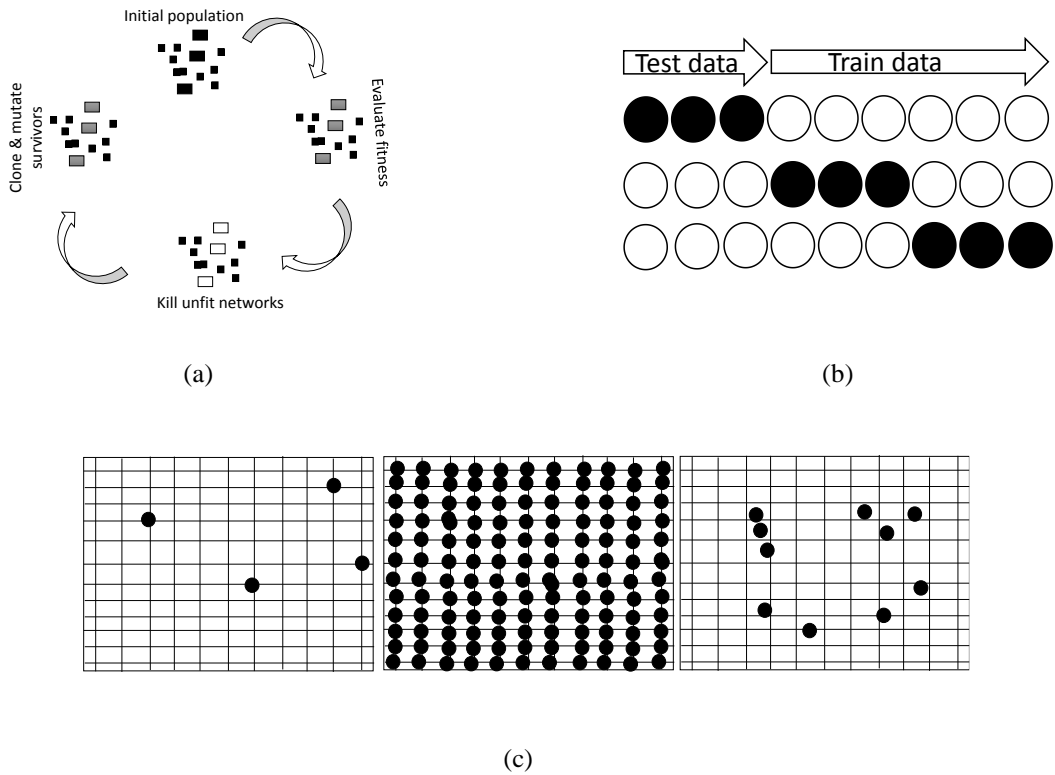

Fig. 7. a. Genetic algorithm, b. k-fold cross validation, c. manual, grid, and random search.

The first issue to be considered in ANN NLE (or any other ANN application) is selecting its hyperparameters, which include number and length of training vector, number of neurons, and layer type [39]. However, these hyperparameters should be adjusted precisely, in order to avoid overfitting (Fig.9.c), and have a better performance. The layer type is completely dependent on the linearity or nonlinearity of the input data. For example, RBN (Fig.9.a) [38] is proper for linear data input, or MLP (Fig.9.b) [40] is proper for nonlinear data. For equalizing a channel with unstable parameters common ANN is not proper; therefore, it is better to use adaptive ANN (e.g., bi-dimensional equalization with backpropagation MLP uses two MLPs to equalize in phase and quadrature of modulated signal to implement an adaptive equalizer). Some adaptive ANNs use an activation function with a variable $\alpha$ parameter; it is proved that this technique has a faster convergence and results in better equalization [41]. The inter-band cross phase modulations between users [42] as well as inter-user interference [43] could severely degrade the performance of multi-user OC systems. However, it is shown that a complex valued multi-level ANN could solve this issue [42], and has better performance compared with the least mean square (LMS) [43], and IVSTF [44].

The second important ANN implementation scenario is something that is not investigated in SVM at all, the optical performance monitoring. Estimation maximization of the optical link capacity can be obtained in weak nonlinearity, which requires link conditions, e.g. the linear and nonlinear optical signal to noise ratio (OSNR) [45, 46]. ANN is a good choice even for performance monitoring of OC systems, without requirement to prior information of the link. Because, ANN is able to model any relationship [47]. The results of performance monitoring could be used in designing optical system and device. One of the most important things that could be seen in papers of this subject is the wide range of the feature extraction techniques. Actually, extracting features is one of the most important things in ML. Higher order modulations greatly suffer from NLPN, so one of the choices for feature extraction could be the NLPN itself, or second-order statistical moments. This method has high accuracy, reduces provisioning margins, and increases resource utilization, and due to its quick response (in OSNR monitoring) is a good choice in real-time applications. The other choices (with high dynamic range and accuracy) for extracting features include the eye diagram [47-49], delay-tap plots [50], asynchronously sampled signal amplitudes [51], asynchronous amplitude histograms (AAH) [52], and asynchronous diagrams [53] (in joint monitoring of OSNR, chromatic dispersion (CD), and polarization mode dispersion (PMD)). Multi-parameter monitoring based on these features is cheap, and does not need synchronous sampling, clock or timing recovery. It is shown that asynchronous diagrams of balanced detection have better results than single-ended detection [53, 54]. Features of AHHs are also effective for modulation format identification (MFI) in heterogeneous Fiber OC networks [55]. Most of the presented works in joint monitoring are inn ANN technique; however, [56] used ANN, k-NN, and decision tree to process wavelengths, OSNRs, and bandwidths of optical signals based on the spectral data. For the need of the practical application, it also investigated variable wavelength, OSNR, and bandwidth.

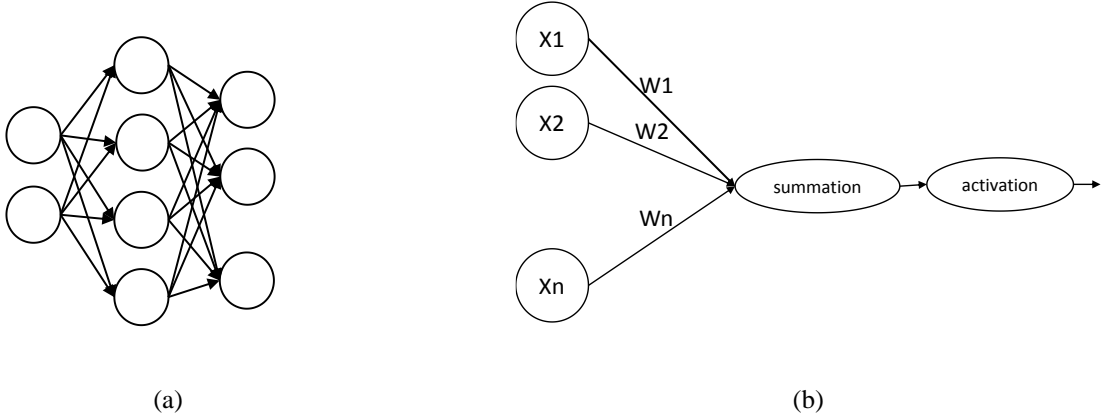

*Fig. 8. a. an ANN, and b. computation process of a neuron.*

Generally speaking, ANN has more applications than SVM (because it has nonlinear nature); one of the other applications of ANN is adjusting the optical amplification gain. The adjustment of the amplification gain is very important, because it can cause ASE noise and nonlinearity. Solving this issue using the adaptive adjustment techniques becomes somehow hard while considering some constraints on this problem, such as power masks makes problem. Whereas using MLP (because of non-linear nature of this problem) could well solve it, and smoothly adjust the gain of each amplifier [57].

**Network:** Now, its turn to review the optical networks applications of ANN; these works are completely Fiber OC based, future works may investigate the ML for WOC networks. Modern OC networks are served as the backbone of cloud-based platforms and are encountered with increase in traffic and reliability requirements. One of their challenges is the cost-effectively of cognition driven learning and fault management techniques. ANN is appropriate for addressing network assurance via dynamic data-driven operation, as opposed to static pre-engineered solutions [58]. Monitoring and modeling of Fiber nonlinearity are important in elastic optical networks (EON)s. However, conventional methods do not provide high accuracy for this task. As another drawback, multi-domain EONs with alien wavelengths, the intra-domain and inter-domain alien traffic should be monitored for estimating QoT of each lightpath and applying provisioning operations. These issue are considered in the existing literature in ML for OC, e.g. [59] used a dual-functional ANN to first calibrate the deviations of the existing models, and to combine link modeling and monitoring for better estimating variance of fiber nonlinear noise. Furthermore, [60] developed an alien wavelength performance monitoring technique and ANN-based lightpath QoT estimation. An ANN-based OSNR monitoring of an OC network is presented in [61], which in a Software Defined Network, a monitoring database decides configuring a probabilistic shaping based bandwidth variable transmitter and increases capacity by adapting the spectral efficiency. It is proven that use of ANN for selecting optimal symbol rate and modulation format in bandwidth variable transceivers in meshed optical networks with cascaded ROADM filtering at fixed channel spacing has higher average capacity compared with a fixed symbol rate transceiver with standard QAMs [62]. As a final note, a comprehensive investigation is done in [63], and various ML techniques, including ANN, SVM, logistic regression, K-nearest neighbors, decision tree and Naive Bayesian are compared in applications such as attack detection accuracy, true positive rate and true negative rate. Further, it is proven that ANN is the most accurate in detecting out of band power jamming attacks in optical networks. ANN is able to predict network traffic [64], and reduces latency as well as learning uplink latency and achieving flexible bandwidth allocation [65]. Using ANN, the flexible bandwidth allocations can be achieved, which leads to low-latency communication.

**WOC:** Previous paragraphs focused on Fiber OC, this paragraph focuses on the applications of ANN in WOC systems including FSO, and VLC (which include few works). The most important issue in front of FSO system is the turbulence of the propagation media. The sensor-less adaptive optics could be used for compensating wavefront disturbance of FSO signal. Use of BP-ANN (Fig.10.a) for this task requires few online measurements compared with other model-based approaches; this is good for real-time applications [66]. This was the only use of ANN in FSO, other works focused on VLC. By using White Light Emitting Diode (LED) in VLC could provide illumination and data transmission simultaneously [67, 68]. The most important issues of VLC include multipath-induced ISI [67, 69], and fluorescent light interference (FLI) [68], which cause high optical power penalty. The ANN could mitigate the ISI (when the transmission rate exceeds the system modulation bandwidth). Furthermore, it is shown that to in the presence of FLI and ISI, use of discrete wavelet transform and ANN based receiver has better performance compared with conventional methods of employing a high-pass filter and a finite impulse response equalizer [67, 68]. The same as single-input single-output (SISO) structure, ANN could also be used as equalizer in a multi-input multi-output (MIMO)-VLC system [70]. Filtering the blue component of the LED (at the cost of the power contribution of the yellowish wavelengths) is a popular method for data rate enhancement. However, with ANN equalizer, it is possible to obtain higher data rates from white light than blue component due to the obtained high SNR from maintaining the yellowish wavelengths [68].

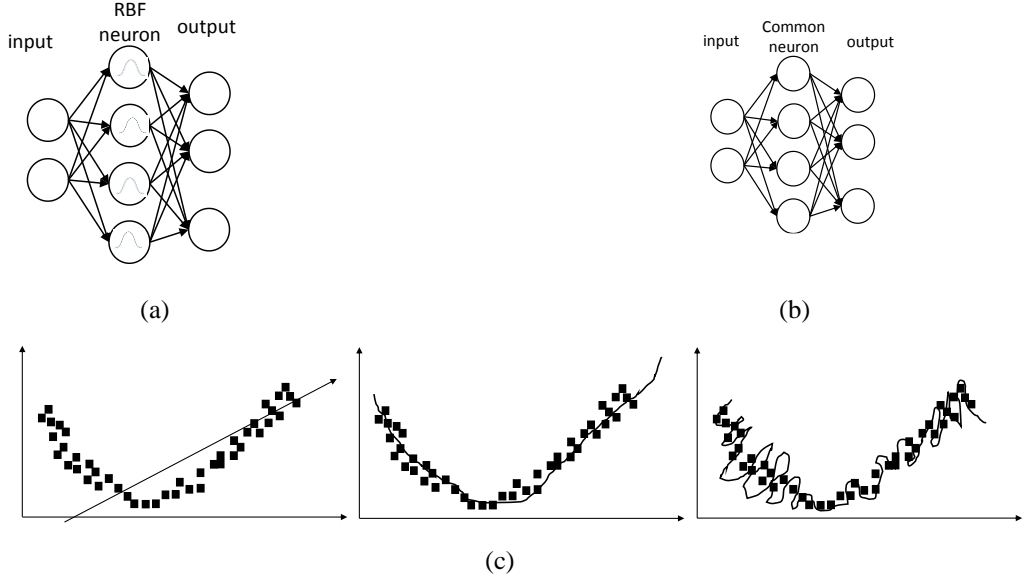

*Fig. 9. a. RBF NN, b. MLP NN, and c. Underfitting (left), good fitting (center), and overfitting(right).*

Extreme learning machine (Fig.10.b) (ELM) is a feedforward ANN and can be used for classification, regression, clustering, sparse approximation, with a single layer or multiple hidden layers. The parameters of hidden nodes need not be tuned, and are randomly assigned and never updated, or can be inherited from their ancestors without being changed. Often, the output weights of hidden layers are learned in one step that essentially amounts to learning a linear model. Despite its good performance and speed, ELM in OC is only applied as a VLC MIMO detector in the presence of LED nonlinearity and cross-LED interference [71]. Most of ANN investigations in OC considered common NN layer types; however, ANN is not a fixed thing, the hyper parameters of ANN could be tuned manually [64] or greedy; in addition to these techniques, hyperparameters could be variable, for example number of neurons could be variable and an algorithm could be deployed to adaptively change it, this is ELM, and could be favorable for applications that no information is available about the data [72].

### C. Deep Neuronal Network

**Definition:** Deep learning is a branch of ML that uses different types of NNs, and is becoming especially exciting now, because there are more amounts of data. Deep learning includes different structures such as DNN, deep belief networks, recurrent neural networks and convolutional neural networks (Fig.11. a, b, c, d). Each of these techniques has success in some specific applications (they are data dependent), and can't compare them unless being used on the same data set. DNN is an extension of ANN that could derive more complex relationships between the input and output by use of multiple additional hidden layers.

**Fiber OC:** Exactly the same as ANN, the main application of DNN is equalization, but because DNN could find more complex relationships, it could be used for more complex channel/system models. In OC, the Shannon capacity (the fundamental barrier of the maximum achievable rate) is mainly attributed to linear and nonlinear distortions [73], fiber Kerr effect [74], and ASE noise. Earlier impractical approaches compensated only deterministic nonlinearities, however, DNN tackles the interplay of deterministic and stochastic nonlinearity manifestation [75]. Furthermore, DNN could compensate nonlinear inter-carrier crosstalk effects even in the presence of frequency stochastic variations. Non-orthogonal waveform signals could improve spectral efficiency, but they cause interference. To solve this issue, could use the spectrally efficient frequency division multiplexing, but it is a complex approach. DNN could be well used to compensate linear and nonlinear interference with low complexity [73]. Another approach for this task is using DNN based on unrolling the split-step Fourier method (SSFM) [76]. This technique (DNN based Digital Back Propagation (DBP)) significantly reduces the complexity compared to conventional DBP, and compared to standard "black-box" DNN. DNN moves through the layers calculates the probability of each output. For example, in symbol detection, DNN calculates the probability that a specific symbol is received. Due to improvements in DNN compared with ANN, this idea could be extended from SISO (in ANN) to MIMO (in DNN) systems, and DNN is applied in scenarios such as Mode Division Multiplexing (MDM) [77], and Spatial Division Multiplexing (SDM) [78]. For example [77] used DNN to implement a somehow inverse channel matrix and use the result for detection, or [78] used Elman NN (ENN) (Fig.12.a) for traffic prediction in an EON. ENN is a typical ANN, with additional layer called context layer, which is connected to the hidden layer with a weight of one, and are used to save previous values of the hidden layer.

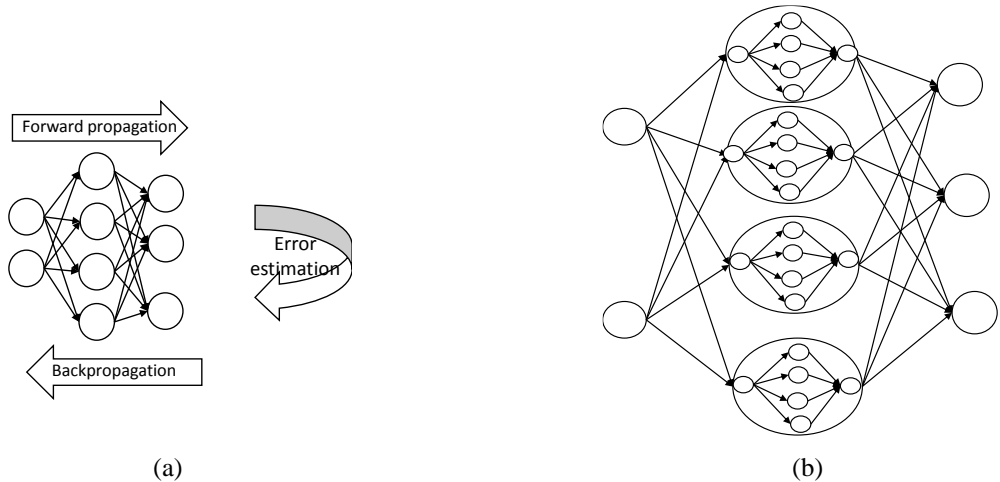

*Fig. 10. a. Back propagation NN, b. ELM NN.*

The second application of DNN is joint monitoring and identification. Again, the same as ANN, the amplitude histograms (AH)s, obtained after the constant modulus algorithm (CMA) equalization, could be used as feature for MFI [79], or joint MFI and OSNR monitoring by DNN [80]. DNN is simple and low cost because it is non-data-aided and avoids any additional hardware on top of the standard digital coherent receiver. In addition MFI using DNN has higher accuracy compared with ANN.

The third application is end to end DNN, which has emerged in OC recently. Mostly, DNN is used as a tool at the receiver side; however, recently, some works investigated the effect of using it at both sides of transreceiver. Actually, end to end DNN enables joint optimization of OC system transceiver parameters. All elements of the transceiver and the fiber channel could be modeled, and by DNN the transceiver configuration could be adjusted by minimizing a loss function. This technique could be used for auto encoding [81-83] or constellation shaping [84, 85], and power adjustment [86] in OC systems. The method is used as is without any further analytic derivations necessary since the ML optimization method is agnostic to the embedded channel model. Long short term memory (LSTM) RNN (Fig.13.a) is one type of RNN, which is composed of a cell, an input gate, an output gate and a forget gate. The cell saves values over arbitrary time intervals and the gates regulate data flow into and out of the cell. OSNR monitoring is important for achieving a reliable and reconfigurable network with high QoS. The use of LSTM-RNN is useful for identifying the OSNR relationship of time-varied data with high accuracy and short response time [87].

The forth and interesting application of DNN is modelling system parts, e.g., optical amplifier [88], and optical modulator [81]. For doing this task DNN learns the relationship between features and targets (e.g. Raman gain profile and pump powers and wavelengths). It is proven that modeling devices using DNN has high accuracy, low latency, and low complexity. The most important parameter in modeling, is the DNN layer type, which is commonly feedforward layer (Fig.12.b), in which data flows in forward direction. However, it could not model the channel memory effect, and for this task, recurrent layer should be used.

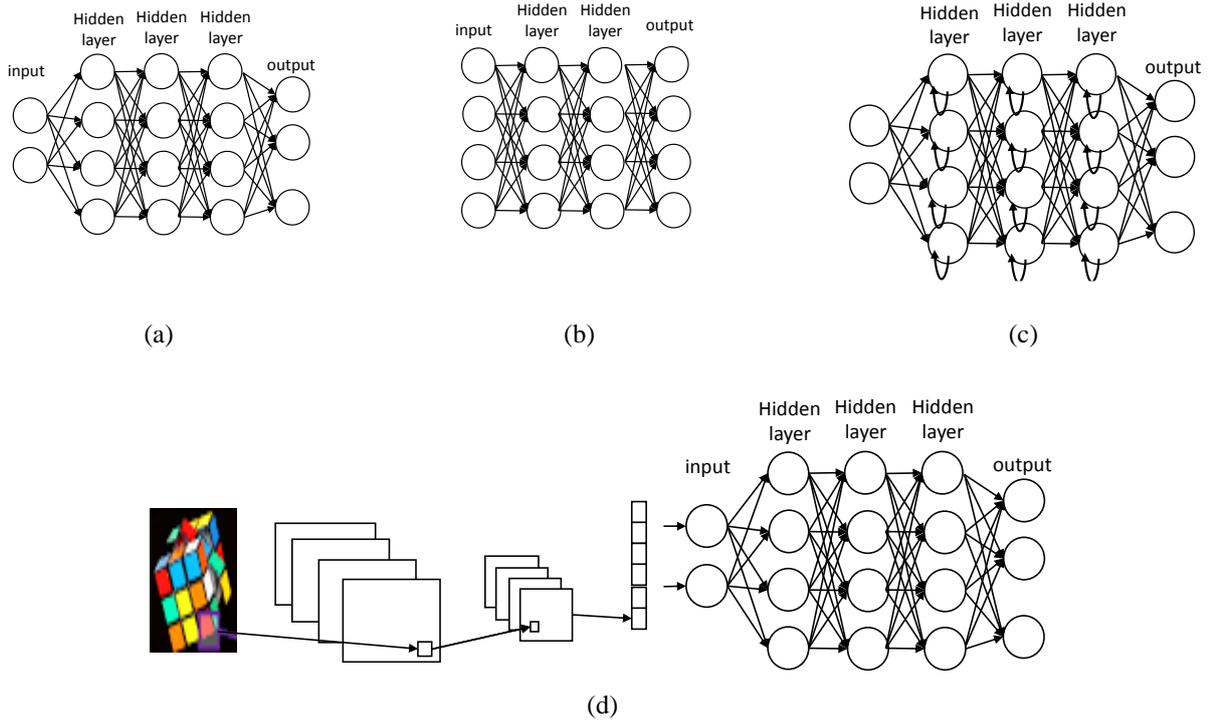

*Fig.11. a. DNN, b. DBN, c. RNN, and d. CNN.*

**Network:** The fifth application of DNN is in optical networks, which is only limited to one investigation. Bidirectional LSTM is designed for combating drawbacks of MLP, and time delay NN (TDNN) (Fig.13.c), on the input data flexibility (MLP, and TDNN have fixed input data), and RNN, on the future input data (RNN cannot achieve future data). Bidirectional LSTM-RNN increase the amount of input data by connecting two opposite directional hidden layers (backward and forward states) to the same output (Fig.13.b). [89] proposed a multi-domain routing paradigm that transforms the inter-domain routing problem from heuristic-algorithm-based computation to AI-based data analytics.

### D. Convolutional Neuronal Network

**Definition:** MLP is a fully connected NN, which is prone to over fitting; for solving this problem, addition of a regularization term related to NN weights is suggested. CNN is a class of DNN (a regularized version of MLP), which is mostly used in analyzing image signals, which presents a different regularization approach. CNN considers hierarchical pattern in data, and assembles more complex patterns using smaller and simpler patterns. The basic idea of connectivity pattern between neurons in CNN is taken from resembles the organization of the animal visual cortex. Different cortical neurons have limited different stimulation areas (receptive fields) that are partially overlapped and cover the entire visual field. CNN is mostly used for image processing. Due to its high computation, few works considered this technique, and other NNs are preferred to it.

**Fiber OC:** Despite previous ML algorithms, the main application of CNN is optical performance monitoring/MFI. By using multi-layer self-learning, CNN can discover intrinsic features of the original image, identify the image, and achieve great success in the OC applications such as optical performance monitoring and MFI [90]. CNN can process constellation diagram as raw data (i.e., pixel points of an image) from the perspective of image processing, without manual intervention nor data statistics. Therefore, eye diagrams or AHHs could be used as inputs of CNN for OSNR monitoring and MFI [91]; this approach outperformed decision tree, KNN, SVM, and ANN. The MFI is indispensable for carrier phase recovery in a coherent optical receiver. Because constellation diagrams of modulation signals are susceptible to various noises; in such cases, a good idea is use of hybrid ML technique, in the sense that first a ML MFI based clock recovery could be developed, and then a CNN could be utilized to process the amplitude data [92].

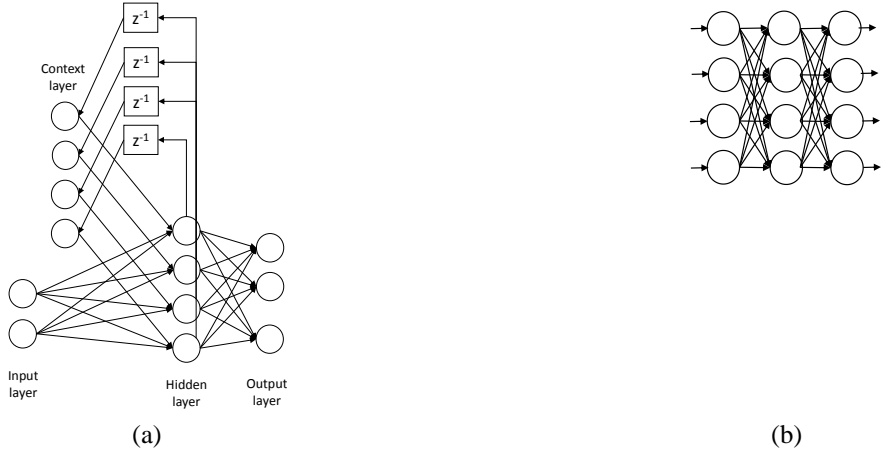

Fig.12. a. Elman NN, and b. feedforward NN.

**Network:** Considering CNN applications in OC networks, they include identification/estimation of target quantities, OPM [93], and QoT estimation [94]. Despite other ML algorithms (used for OPM, and QoT estimation), CNN deals with the abundance of training data required for convergence and pre-processing of input data by human engineers needed for feature extraction. CNN could also be used for detecting Fiber link failures (to which the Fiber speckle gram sensors are highly sensitive) [95]. At first different images are taken from perturbed fibers, then the CNN is learned, then a new image is checked for perturbations. This process also could be done by ANN, but a bit change in input features is required, because they are fundamentally different.

**WOC:** Most of ML for FSO investigations considered CNN. CNN could be used a as a demodulator for a turbo-coded FSO system in the presence of strong atmospheric turbulence [92]. The feasibility of the scheme is verified by transmitting a two dimensional digital image. CNN also could be used for joint atmospheric turbulence detection and demodulation for an orbital angular momentum FSO system [90, 96]. It is proven that this technique has higher accuracy compared with previous approaches. Although vortex beam carrying orbital angular momentum increases capacity, atmospheric turbulence affect it by distorting the helical phase fronts of the propagating beams. Recently, CNN has been designed to combat this effect by feature extraction from the received Laguerre–Gaussian beam's intensity distributions; in the proposed structure, there is a tradeoff between the computational complexity and the efficiency of achieving a good recognition of OAM mode in atmospheric turbulent [97]. As the only investigation of CNN in VLC system [98], CNN is used for demodulation. CNN is the classifier of imagery data input, accordingly it converts the modulated signals to images and recognizes them by classifying these images.

### E. K Nearest Neighbor

**Definition:** The k Nearest Neighbor (kNN) is a non-parametric ML method (Fig.14.), which classifies the input based on a majority vote of the k closest training neighbors in the feature space. For example, when $k = 1$, the input is assigned to the nearest neighbor class. The kNN calculates the Euclidean distance from the input point to all of the training set points by

$$D(R, L_i) = \sqrt{(R - L_i)^2} \tag{8}$$

where, $R$ is the input point, $L_i$ is the training data. Then it sorts all of $D(R, L_i)$ to take k nearest neighbors, the estimated neighbor is classified by a majority vote of its k nearest neighbors [99]. In kNN, the function is approximated locally and all computation is deferred until classification. The kNN detection is the same as soft detections such as maximum likelihood based on the probability density function of the received signal [100]. Assigning weights to the contributions of the neighbors is one way to improve performance.

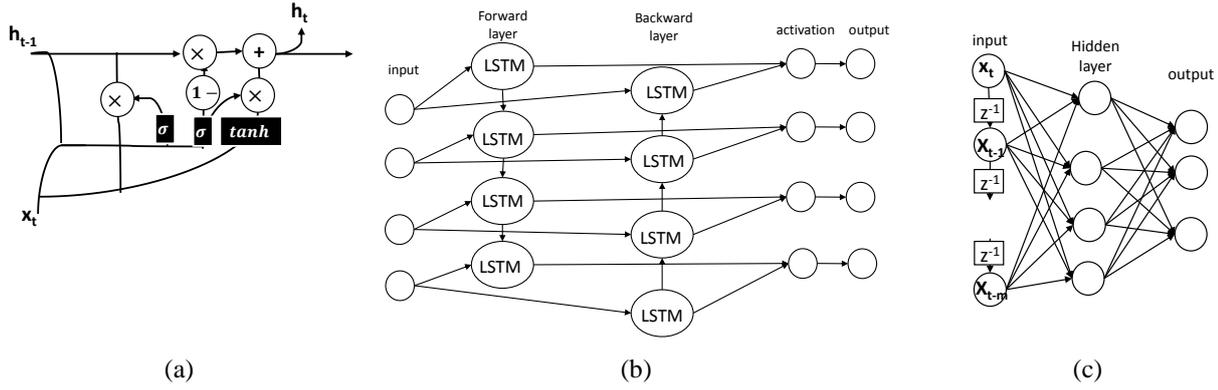

*Fig.13. a. LSTM-RNN, b. Bidirectional LSTM-RNN, and c. TDNN.*

**Fiber OC:** The same as SVM, kNN is mostly used as detector in literature, and could deal with effects such as NLPN, ASE noise, SPM. NLPN could be compensated partially by mid-span optical phase conjugation (OPC). kNN detector is effective for the circularly symmetric non-Gaussian impairments, such as ASE noise, I/Q imbalance, and NLPN of dispersion unmanaged link, and could be used to further improving performance of the OPC by mitigating non-Gaussian impairments caused by NLPN [100]. However, one of the problems of kNN is the distributed training points might not be uniformly distributed. In this situation the nearest class might have one point, and the farther class might have more points, and finally the input point would be dedicated to the farther class because of more points. As a solution for this problem, Distance Weighted kNN (DW-kNN) detector emerged, which gives a weight related to the distance to each of the surrounding points and then takes a decision; this technique better mitigates fiber impairments, and outperform maximum likelihood post compensation approach [101]. However, the kNN and DW-kNN still have some complexity, because they are dependent on the amount of input data and the value of k. As a solution to this problem, the non-data-aided kNN (ND-kNN) provides efficient nonlinearity mitigation with low complexity and zero data redundancy [102]. The ND-kNN compensates any non-deterministic impairments and does not require any extra training data. It uses the density parameter of the test data to rapidly extract the center noise-less data and label them as the classification references; then applies the kNN to classify the remaining test data. Therefore, it is robust to noise and has fast convergence as well as significant BER improvement in back-to-back and SMF transmission.

### F. Random Forest

**Definition:** When the input point of estimated point belongs to the insensitive cluster, the KNN could not determine the neighborhood of the detection point, so a method that does not rely on distance calculation should be used. Random forest algorithm is a bagging algorithm based on decision tree model, a set of decision trees $DT = \{dt_1 \, dt_2 \cdots dt_n\}$. Random forest is an ML technique that develops training based on the constructed decision trees (forests); its output is the mode of the trees output classes (in classification) or the mean of the trees output predictions (in regression) (Fig. 15.).

**Network:** Nowadays, considering convenience of ANN, and SVM, other ML techniques are not so much considered in ML for OC applications. However, some decision trees Classifier ML algorithms such as Ada Boost, Decision Tree, Random Forest, and Gradient Boosting. Random forest could be used for estimation of BER of an unestablished link or QoT [103, 104], and estimation of received optical power parameter [105]. Random Forest does not have any parameter to be tuned and its performance is completely related to the structure of the presented trees and forest.

### G. Regression

**Definition**: Regression is composed of statistical processes for deriving the relationships between variables, and includes several techniques. Actually, it helps better understanding changes of a dependent variable as a response to the variation of one of the independent variables, while fixing other variables. Regression estimates regression function (a function of the independent variables); mostly, it estimates the average of dependent variable while fixing independent variables; sometimes, the focus is on a quantile, or other location parameter of the distribution of the dependent variable while fixing the independent variables. Regression analysis is widely used for prediction and forecasting, where its use has substantial overlap with the field of machine learning.

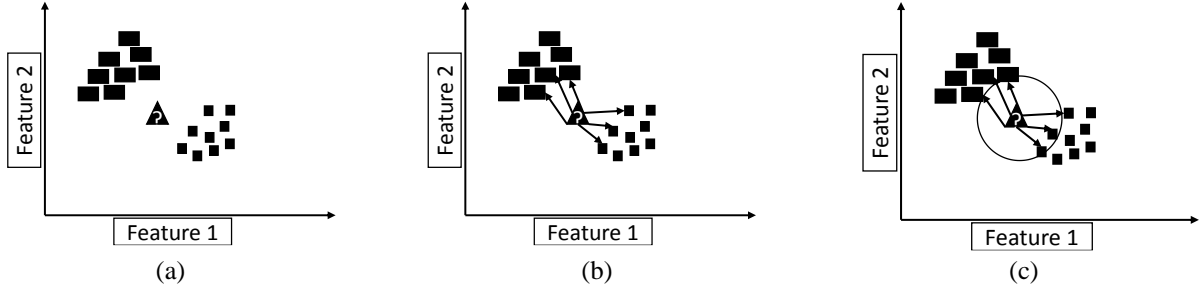

*Fig.14. kNN. a. initial data, b. calculate distance, and c. finding neighbors and voting for labels.*

**Fiber OC:** Least Absolute Shrinkage and Selection Operator (Lasso) regression (Fig.16.a) is a linear regression algorithm, which shrinks data towards a central point, and develops simple, sparse models. In order to identify the interaction between symbols in different time slots and to select the minimum number of relevant employed perturbation terms, sparsity phenomena is required. Considering this scenario, Lasso regression is a good choice for simplifying nonlinearity mitigation analysis without losing the important key features [106, 107].

Supervised algorithms have a loss function that should be minimized; the Gradient descent is a well-known algorithm for this minimization. It could further extended and used as a learning algorithm, and be used in OC applications (e.g. QoT [108]). Although fiber optic mathematical modeling is hard, some papers deployed some simplified OC models for ML. However, it should be notified that ML has reduced uncertainties on input parameters of these models, by improving the accuracy of the SNR estimation based on a propagation model.

**Network:** Maintaining channel power stability during fast-changing spectrum utilization is crucial to ensuring the QoS in flex grid OC networks. These networks require spectrum defragmentation to improve spectral efficiency. Various defragmentation methods, e.g., hop, make before break, sweep, and interact with the power dynamics of EDFA differently. EDFA power excursions is a challenge for network operations and can result in exacerbated post-EDFA power discrepancy during spectral defragmentation processes. ML could be used to accurately predict the power excursions in WDM channel provisioning, leads to quick and precise decisions addressing network demands, and optimizes EDFA power dynamics. Lasso regression (Fig.16.b) adds an L1 regularization term to the cost function, and result in sparse models with few coefficients. If someone wants not to eliminate weights, but reduce non-important ones, use of larger regularization term is a good choice. Ridge regression is a linear regression with additional L2 regularization term. In Kernelized Bayesian regression (Fig.16.c), first input features are mapped to another higher dimension, then regression model determines the posterior pdf of it. These techniques in a fully automated workflow extract EDFA power dynamics and adjust power without iterative power measurements [109- 110].

**WOC:** The only investigated application of regression in VLC is visible light positioning (VLP). A rarely investigated ML structure is the dual-function ML, which could be used applications such as Radio Transformer Network, indoor positioning, etc. This technique combines two ML methods which each of them are proper for a single task. For example, in indoor positioning, different indoor areas could be divided into some parts by a classification algorithm, and then a regression algorithm could be used for exact positioning [112].

### III. Unsupervised Learning

In unsupervised ML data is unlabeled, and the algorithm tries to find patterns and structures in data. Accordingly, important key features are kept and a new dataset with reduced redundancy is produced. For example in link failure detection, algorithm finds that there are two different categories, which each of them has something in common. Actually, unsupervised learning works within the data without reference given from the outside [113]. Unsupervised learning is a type of ML algorithm used to draw inferences from datasets consisting of input data without labeled responses.

#### A. Peak search, c-means, hieratical clustering

**Definition:** In spite of supervised algorithms which are famous by their names (e.g. SVM, ANN, etc.), clustering algorithms, are mostly defined under the shelf of clustering, and mostly the programmer decides the algorithm what to do (based on some logical rules). For example, in non-iterative density peak search clustering algorithm (Fig.17.c) [114] two simple parameters are used for seeking the cluster centers without any iteration, in contrast Fuzzy c-means does the same thing iteratively; in opposite to both of these techniques, Hieratical clustering solves the problem blindly [115]. In another clustering algorithm, called fractionally spaced clustering, Viterbi type procedure along with Mahalanobis distance metric is used, dictated by the specific transitions among clusters [116]. So, the flexibility of unsupervised (clustering) algorithms is high, and in addition they do not require label, which makes them really helpful. However, they have one single aim; find the center and grow up the cluster.

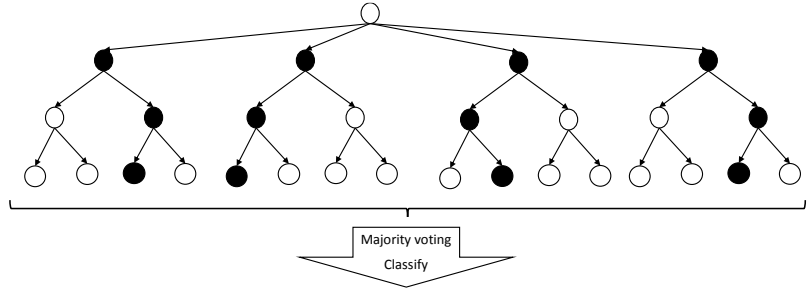

*Fig.15. Random forest*

**Fiber OC:** The k-means (Fig.18.a) and Principle Component Analysis are the mostly used unsupervised algorithms in OC; however, there are some other techniques, which could not lie in a major group e.g. Hierarchical (Fig.17.a) and Fuzzy-logic C-means (Fig.17.b) clustering. These techniques try to find hidden patterns or groups in data, and have lower complexity than supervised algorithms; however, their drawback is taking more convergence time. In spite of the generality and wideness of these techniques, they have done well compared with conventional methods in MFI [114, 115], NLE [116], etc. Furthermore, in contrast to traditional algorithms (e.g. deterministic Volterra NLE), these techniques could combat stochastic parametric noise amplification, and reduce complexity and processing time by deploying blindly [20].

A. **k-means clustering**

**Definition:** The k-means partitions data into k clusters, and assigns each data point to the nearest mean cluster. This problem has high computations; however, heuristic algorithms converging to local optimum can be used. The "k" refers to the number of clusters, which should be identified among *L* observations, such that the sum of the squared error of data points be minimized. At first this algorithm takes randomly K center cluster positions. Then each data point is assigned to the cluster with the nearest center, then based on the defined clusters, a new cluster center is defined (the summation of the cluster points). The same procedure is repeated until convergence.

**Fiber OC:** Time-varying distortions, derived from I/Q imbalance, NLPN, and inter-channel interference, result in centroid deviation of received symbols. Employing short time windows for detection enables inline tracking of signal variations based on centroids and cluster deviations. These issue could be solved by blending the time windowing phenomena and k-means detection [117]. To reduce the complexity and redundancy, and to improve speed and accuracy of the standard k-means other k-means algorithms such as training sequence assisted k-means, as well as blind k-means are proposed. In [118] a traditional Density-Based Spatial Clustering of Applications with Noise (DBSCAN) (Fig.18.b) NLE is presented. Furthermore, this algorithm is combined by k-means clustering and implemented on the noisy "un-clustered" symbols. In [119] a Parzen window classifier (Fig.18.c) based detector is used to mitigate the fiber non-Gaussian nonlinear effects in dispersion managed and unmanaged fibers, by designing improved decision boundaries. Parzen window algorithm is a non-parametric method of estimating the class-conditional densities (likelihoods) in a supervised pattern classification problem from training data. The proposed k-means algorithms proved to be proper for mitigating nonlinearity effects [120, 121], and parametric noise of cascaded EDFA [122].

Few RoF systems are investigated in ML for OC applications; however, they mostly used k-means based detector to mitigate fiber effects [121-123]. The k-means is simple and a good choice for RF phase recovery (for phase modulation) [122], and it is a good RF carrier phase method for higher order modulation formats [123]. By grouping highly correlated neighboring samples into multi-dimensional vectors and adopting k-means clustering for quantization in digitized RoF systems, [121] demonstrated spectrally efficient digitized RoF transmissions. This technique has better performance than PCM based RoF system.

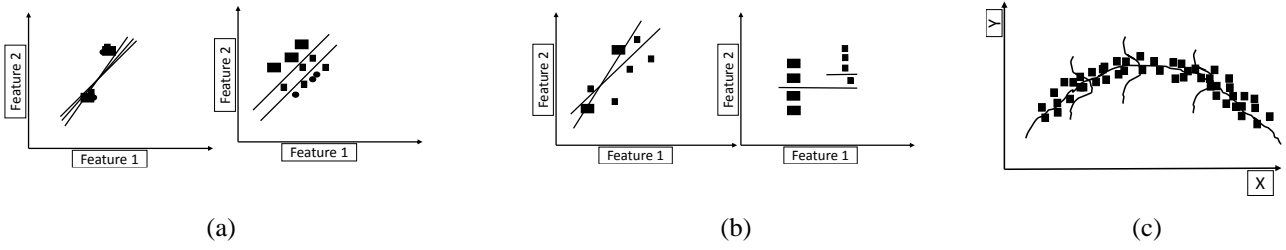

*Fig.16.a. Lasso, b. Ridge, and c. Bayesian Regression*

**Network:** The only instigation of k-means in OC networks is the deployment of green field passive networks to minimize the total deployment cost [124].

**WOC:** by increase in human underwater activities, underwater VLC attracted many attentions toward itself. K-means algorithm, despite less complexity, could deal with noise and nonlinearity, as well as amplitude and phase jitter (as the main drawbacks of VLC system). For example, [125] proposed a k-means clustering technique to correct the phase deviation of special-shaped 8-QAM constellations. As another example, [126] proposed a time-amplitude two-dimensional re-estimation based on DBSCAN algorithm to distinguish different signal levels with jitter. DBSCAN groups together a set of points of a space (high density region), and marks as outlier the points that are in low-density regions.

### B. Expectation Maximization

**Definition:** The drawback of K-means clustering is its hard decision boundaries that assign data point only to one cluster; however, it might lie between two or more clusters. The EM clustering assigns a probability of belonging to a cluster to the data point rather than completely assigning it to a particular cluster. This technique assumes that the data points have a superposition of K jointly Gaussian probability distributions with different means and covariance (Gaussian mixture model) (GMM) (Fig. 18.d). Actually, in both of them, n iterative refinement approach along with cluster centers are used to model the data; however, clusters of k-means have comparable spatial extent, but EM clusters have different shapes.

**Fiber OC:** Bayesian filtering in combination with EM is used for mitigating the laser amplitude and phase noise [127], as well as nonlinearity compensation, carrier recovery, and nano-scale device characterization [128]. These technique use Bayesian analysis and evaluate the content of an incoming optical signal and determine the probability that it constitutes amplitude or phase noise.

### C. Principle Component Analysis

**Definition:** PCA is an orthogonal transformation of possibly correlated data into a linearly uncorrelated data entitled principal components. The first component of PCA has higher variance, and each succeeding component in turn has the higher variance and is orthogonal to the preceding components. This transformation could be done by eigenvalue decomposition of a data covariance matrix or singular value decomposition of a data matrix. Considering a set of input data as $\{s_1, s_2, \ldots, s_N\}$, $N$ vector of $M$ dimensions, the mean and covariance vectors will be calculates as:

$$\bar{x} = \frac{1}{N}\sum_{i=1}^{N} x_i, \tag{10}$$

$$\Sigma \approx \frac{1}{N}\sum_{i=1}^{N}(x_i - \bar{x})(x_i - \bar{x})^T, \tag{11}$$

where $\Sigma$ has $M$ eigenvalue ($\lambda_i$) and eigenvectors ($\mu_i$). PCA sorts eigenvalues, and selects the first $P$ eigenvalue ($P \ll M$).

**Fiber OC:** PCA is mostly suitable for feature extraction; (considering discussions of previous sections) its main application is in joint monitoring of OSNR, CD, and PMD of different modulation formats and data rates [129]. It is proven that PCA could accurately identify the data rates and modulation formats of the signals in the presence of noise, CD, and PMD.

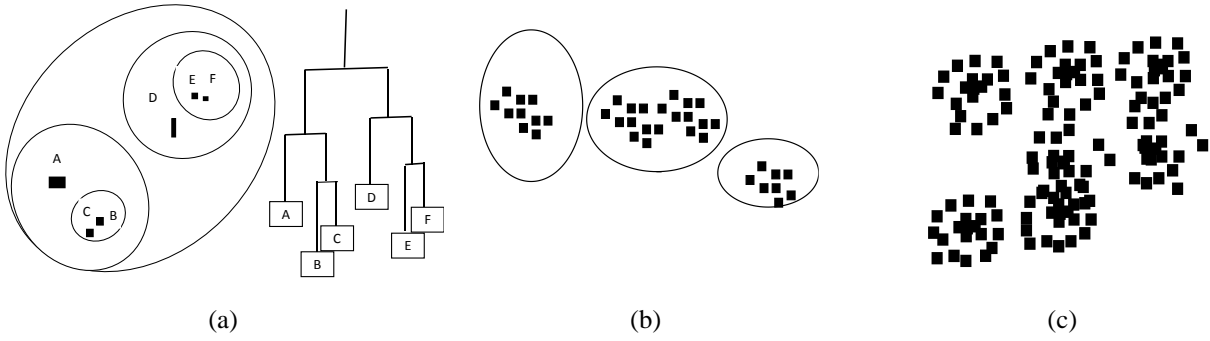

*Fig.17. a. Hierarchical clustering, b. Fuzzy-logic C-means clustering, c. peak search clustering.*

**Network:** The PCA could be used for arranging priority, and recognizing abnormal or attack flow [130]. The abnormal attack flow changes the correlation between the flows; PCA extracts the correlation between potential abnormal parts of multiple flows and uses the degree of correlation changes as a measure [131]. Optical spectrum (OS) has an important role in optical signal processing, such that ultra-high resolution (UHR)-OS provides more and accurate information about optical link quality. Comparing experimental and theoretical UHR-OS is one way for detecting signal distortions. However, the monitored UHR-OS of the channel might not be fixed or known. Also measuring the ideal UHR-OS at or near the transmitter might be impossible. It is shown that use of PCA for feature extraction from observed OS, and use of SVM for classification, could combat these issues and retrieve UHR-OS [132].

### IV.     Reinforcement Learning

**Definition:** Reinforcement learning (Fig.19.a) includes algorithms that learn by rewards for favorite actions. It is based on a Markov decision process, which is a mathematical model for decision making. It is fully dependent on the programmer knowledge and curiosity. The input (observation) is associated with a reward (reinforcement), and the output (action) determines the next input (and accordingly reward). The aim of training is finding the sequence of actions (it tries different actions) that leads to the best reward. So the action is the most important part; the reward and punishment are dedicated to the model (action) when the output is good and bad, respectively. The training here does not mean to learn from a supervisor, or a structure, it means learning from its own experiences.

**Network:** Transparent optical networks (TON)s are the basis of the Next Generation Optical Internet (NGOI); they don't have massive optoelectronic monitoring. However, they have security problems, and require a somehow self-healing to be protected from failures, attacks, and reliability issues [133]. RL has high flexibility, and could be a good choice for increasing reliability, and could generate an intelligent policy to combat failures, and attacks. Optical burst switching (OBS) network is the second promising technology for the NGOI, because it has a relatively easy implementation of optical circuit switching and the efficient bandwidth utilization of Optical Packet Switching (OPS) techniques. The implementation of this method is easy and cost-effective, and does not require either wavelength converters or optical buffers. However, it has high blocking probability, because of wavelength contention. There are several RL techniques considering this issue, for example a deflection routing combined with an adaptive offset time assignment mechanism is used in [134] to deal with wavelength contention and insufficient offset time losses, respectively. Shortest path deflection routing aims to find better deflection paths by use of learning agents at each node to learn the optimal alternative output links at each moment; however, a more efficient RL algorithm for deflection routing is presented in [135]. [136] proposed two RL algorithms for alternative routing, and integrated routing and contention resolution to jointly deal with wavelength issues proactively and reactively. Dynamic path selection has better performance than the deterministic path selection, because it changes by traffic conditions. The single agent path selection, choses a path on the feedback received at the ingress node, and does not affect from selected paths by other network nodes, but multi-agent path selection includes this effect [137].

After OBS, and TON related works, the third main focus is on reducing the blocking probability (or applications that somehow lead to it), by using RL in applications such as quality of service (QOS) estimation, VLC network selection, Designing multicast trees, spectrum sensing in cognitive radio network, routing. Distributed route selection in optical networks subject to physical impairments [138], processing lightpath requests, taking care of lightpath setting up, resource reservation and load balancing [139]. Growing the multimedia applications resulted in increased interest in multicast designing for reducing the blocking probability. It is shown that RL could adapt to network topology and the traffic changes without prior information, so it is a good choice for this application [140]. Another technique that is good for reducing blocking probability is the [141] adaptive online RL based routing, which is based on load shared sequential routing in which RL updates the load sharing factors. The established RL algorithms are either based on automata learning or online learning that are designed for adversarial multi-armed bandit problem. Another technique that might further help reducing the blocking probability is the QoS/QoT estimation. Deserving the QoS requirements is important in OC networks, [142] presented a provisioning strategy for connectivity services with different priorities based on RL able to accommodate QoS requirements and maximize provider benefits. Further, [143] using ML improved the accuracy a set of the estimated QoT data (SNR, power levels, noise figures), and lowered the design margins.

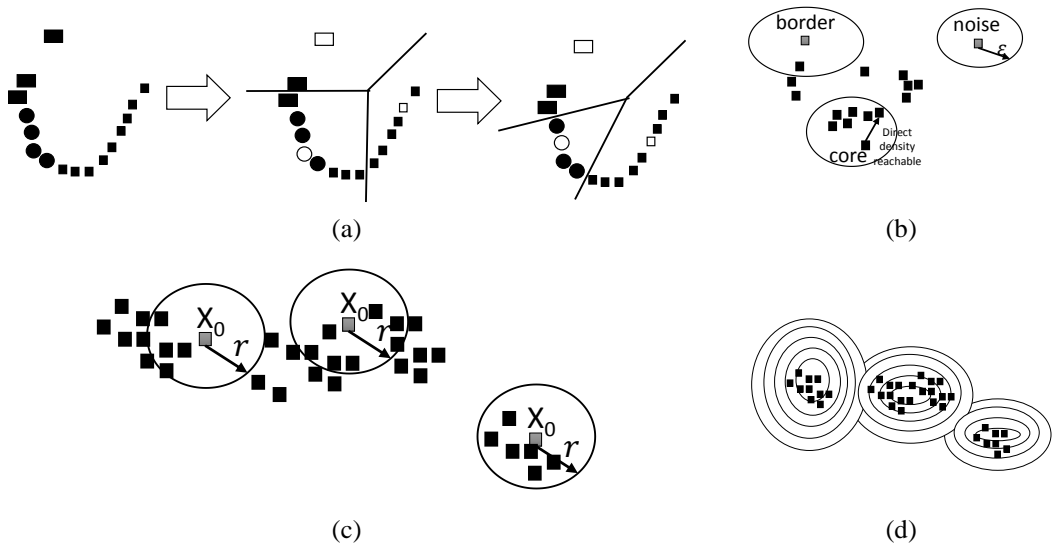

*Fig.18. a. k-means, b. DBSCAN, and c. Parzen window, d. EM GMM.*

The only investigations of WOC networks used RL. One of them used RL in decentralized cognitive radio network, in which secondary users search spectrum opportunities. In this technique autonomous cognitive radios learn distributively their own spectrum sensing policy, so the secondary users non-cooperatively achieve equilibrium, the collisions among them would be reduced, and idle channels could be effectively used [144]. The other one used RL for the network selection in dynamic environments of indoor VLC. [145] presented a context-aware RL algorithm, which uses time location dependent periodic changing rule of load statistical distributions to choose online network via knowledge transfer.

### A. Q-learning

**Definition:** Q-learning (Fig.19.b) is a model free RL algorithm, where "Q" stands for the quality of the taken action of a state. It learns a policy to tell an agent to take what action under what circumstances, and handles problems with stochastic transitions and rewards. Q-learning does not need policy; actually, it learns from random actions that are outside the current policy; it searches for a policy to maximize the total reward.

**Network:** The Q-learning is used in deflection routing to resolve problems of OBS networks [146-148]. In OBS networks, efficient algorithms are very important for choosing path / wavelength that minimizes the burst loss probability; however, it is proved that Q-learning could deal with this issue [148]. At an output node, the path/wavelength selection algorithm calculates the Q values for a set of pre-computed path/wavelength and choses one with the minimum burst loss probability [146]. It is shown that Q-learning could effectively deal with resource allocation problem of Fi-Wi access network supporting IoT service [149]. For this aim, first the dynamics of regular traffic should be analyzed and then Q-learning be used for traffic prediction.

### B. Deep Reinforcement Learning

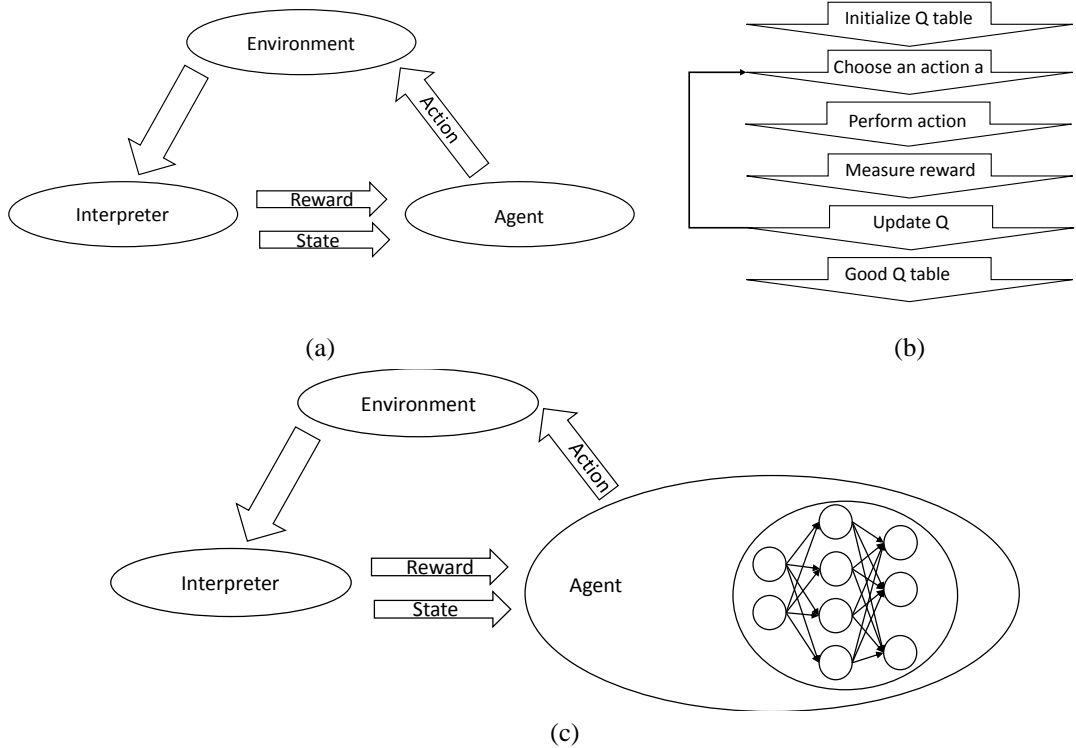

*Fig.19. a. Reinforcement Learning, b. Q-Learning, and c. Deep Reinforcement Learning.*

**Definition:** Deep reinforcement learning (DRL) (Fig.19.c) combines deep learning and reinforcement learning and generates algorithms that can be used in complicated applications that require flexibility e.g., in complicated optical networks. Deploying DL (e.g. DNN) in RL (e.g. Q-learning) generates a powerful tool (DRL) that is capable to scale to unsolvable problems.

**Network:** The flexibility of DRL could also be helpful in EONs, which are highly dynamic. [150] defined a criterion, named the whole network cost-effectiveness value with survivability, and proposed a performance optimization under survivable EON environment. Then a DRL improves performance, in which two agents are proposed to provide working and protection schemes converging toward better survivable routing, modulation level, and spectrum assignment policies. [151] presented a DRL framework for realizing virtual network slicing in inter-datacenter network, in which based on DRL, the infrastructure provider advertises resource and pricing and grants the virtual network embedding schemes.

## V. Conclusion

The number of published papers in Machine Learning for Optical Communication is growing very rapidly. Accordingly, being aware about the last progressions in this field is the most important thing that researchers need. Although there are several works released in this subject, all of them have Optical Communication view. One of the main problems of the researchers in this topic is that they are Optical Communication experts and do not know to use which Machine Learning algorithm, or there is no information about the algorithms used in this topic and many other ambiguities like these. Accordingly, this paper, for the first time, presented a comprehensive overview on this topic with a Machine Learning view. It has worth to mention that the amount of investigated papers in this tutorial is much more than the previously released surveys, which could help the reader to have a good, comprehensive, and reliable information about the released works in this topic.

## References

[1] Bishop, C. M. (2006). Pattern recognition and machine learning. springer.

[2] Wang, D., Zhang, M., Li, Z., Cui, Y., Liu, J., Yang, Y., & Wang, H. (2015, September). Nonlinear decision boundary created by a machine learning-based classifier to mitigate nonlinear phase noise. In 2015 European Conference on Optical Communication (ECOC) (pp. 1-3). IEEE.

[3] Han, Y., Yu, S., Li, M., Yang, J., & Gu, W. (2014). An SVM-based detection for coherent optical APSK systems with nonlinear phase noise. IEEE Photonics Journal, 6(5), 1-10.

[4] Li, M., Yu, S., Yang, J., Chen, Z., Han, Y., & Gu, W. (2013). Nonparameter nonlinear phase noise mitigation by using M-ary support vector machine for coherent optical systems. IEEE Photonics Journal, 5(6), 7800312-7800312.

[5] Wang, D., Zhang, M., Cai, Z., Cui, Y., Li, Z., Han, H., ... & Luo, B. (2016). Combatting nonlinear phase noise in coherent optical systems with an optimized decision processor based on machine learning. Optics Communications, 369, 199-208.

[6] Du, J., Sun, L., Chen, G., & He, Z. (2017, July). Machine learning assisted optical interconnection. In 2017 Opto-Electronics and Communications Conference (OECC) and Photonics Global Conference (PGC) (pp. 1-3). IEEE.

[7] Thrane, J., Wass, J., Piels, M., Diniz, J. C., Jones, R., & Zibar, D. (2017). Machine learning techniques for optical performance monitoring from directly detected PDM-QAM signals. Journal of Lightwave Technology, 35(4), 868-875.

[8] Hui-Ping, Z., Hong-Yan, H., & Meng-Xia, G. (2014, June). Research of Fiber-Optical Fault Diagnosis Based on Support Vector Machine (SVM) Mining. In 2014 Fifth International Conference on Intelligent Systems Design and Engineering Applications (pp. 803-807). IEEE.

[9] Yuan, Y., Zhang, M., Luo, P., Ghassemlooy, Z., Wang, D., Tang, X., & Han, D. (2016, July). SVM detection for superposed pulse amplitude modulation in visible light communications. In 2016 10th International Symposium on Communication Systems, Networks and Digital Signal Processing (CSNDSP) (pp. 1-5). IEEE.

[10] Cui, Y., Zhang, M., Wang, D., Liu, S., Li, Z., & Chang, G. K. (2017). Bit-based support vector machine nonlinear detector for millimeter-wave radio-over-fiber mobile fronthaul systems. Optics express, 25(21), 26186-26197.

[11] Sun, X., Su, S., Huang, Z., Zuo, Z., Guo, X., & Wei, J. (2019). Blind modulation format identification using decision tree twin support vector machine in optical communication system. Optics Communications, 438, 67-77.

[12] Sun, L., Du, J., Chen, G., He, Z., Chen, X., & Reed, G. T. (2017, July). Machine-learning detector based on support vector machine for 122-Gbps multi-CAP optical communication system. In 2017 Opto-Electronics and Communications Conference (OECC) and Photonics Global Conference (PGC) (pp. 1-3). IEEE.

[13] Sun, Y., Hu, J., Tian, Z., Liu, X., & Lu, H. (2017, August). Equalization algorithm based on CMA and SVM for carrierless amplitude phase modulation in optical access networks. In 2017 16th International Conference on Optical Communications and Networks (ICOCN) (pp. 1-3). IEEE.

[14] Liang, A., Yang, C., Zhang, C., Liu, Y., Zhang, F., Zhang, Z., & Li, H. (2018). Experimental study of support vector machine based nonlinear equalizer for VCSEL based optical interconnect. Optics Communications, 427, 641-647.

[15] Wang, C., Du, J., Chen, G., Wang, H., Sun, L., Xu, K., ... & He, Z. (2019). QAM classification methods by SVM machine learning for improved optical interconnection. Optics Communications.

[16] Giacoumidis, E., Mhatli, S., Stephens, M. F., Tsokanos, A., Wei, J., McCarthy, M. E., ... & Ellis, A. D. (2017). Reduction of nonlinear intersubcarrier intermixing in coherent optical OFDM by a fast newton-based support vector machine nonlinear equalizer. Journal of Lightwave Technology, 35(12), 2391-2397.

[17] Giacoumidis, E., Mhatli, S., Le, S. T., Aldaya, I., McCarthy, M. E., Ellis, A. D., & Eggleton, B. J. (2016, September). Nonlinear blind equalization for 16-QAM coherent optical OFDM using support vector machines. In ECOC 2016; 42nd European Conference on Optical Communication (pp. 1-3). VDE.

[18] Nguyen, T., Mhatli, S., Giacoumidis, E., Van Compernolle, L., Wuilpart, M., & Mégret, P. (2016). Fiber nonlinearity equalizer based on support vector classification for coherent optical OFDM. IEEE Photonics Journal, 8(2), 1-9.

[19] Giacoumidis, E., Tsokanos, A., Ghanbarisabagh, M., Mhatli, S., & Barry, L. P. (2018). Unsupervised Support Vector Machines for Nonlinear Blind Equalization in CO-OFDM. IEEE Photonics Technology Letters, 30(12), 1091-1094.

[20] Giacoumidis, E., Matin, A., Wei, J., Doran, N. J., Barry, L. P., & Wang, X. (2018). Blind nonlinearity equalization by machine-learning-based clustering for single-and multichannel coherent optical OFDM. Journal of Lightwave Technology, 36(3), 721-727.

[21] Chen, W., Zhang, J., Gao, M., & Shen, G. (2018). Performance improvement of 64-QAM coherent optical communication system by optimizing symbol decision boundary based on support vector machine. Optics Communications, 410, 1-7.

[22] Ding, G., Wu, Q., Yao, Y. D., Wang, J., & Chen, Y. (2013). Kernel-based learning for statistical signal processing in cognitive radio networks: Theoretical foundations, example applications, and future directions. IEEE Signal Processing Magazine, 30(4), 126-136.

[23] Jarajreh, M. A. (2018). Compensation of filter cascading effects and non-linearities in flexible multi-carrier-based optical networks using a complex-kernel-based support vector machine. IET Communications, 12(14), 1737-1742.

[24] Yuan, Y., Zhang, M., Luo, P., Ghassemlooy, Z., Lang, L., Wang, D., ... & Han, D. (2017). SVM-based detection in visible light communications. Optik, 151, 55-64.

[25] Zhang, J., Chen, W., Gao, M., & Shen, G. (2017, July). Mitigating fiber nonlinearity using support vector machine with genetic algorithm. In Conference on Lasers and Electro-Optics/Pacific Rim (p. s1354). Optical Society of America.

[26] Gu, B., & Sheng, V. S. (2017). A Robust Regularization Path Algorithm for $\nu$-Support Vector Classification. IEEE Transactions on neural networks and learning systems, 28(5), 1241-1248.

[27] Sebald, D. J., & Bucklew, J. A. (2000). Support vector machine techniques for nonlinear equalization. IEEE Transactions on Signal Processing, 48(11), 3217-3226.

[28] Mata, J., de Miguel, I., Durán, R. J., Aguado, J. C., Merayo, N., Ruiz, L., ... & Abril, E. J. (2017, December). A SVM approach for lightpath QoT estimation in optical transport networks. In 2017 IEEE International Conference on Big Data (Big Data) (pp. 4795-4797). IEEE.

[29] Peng, K., Zhang, M., Li, Q., Lv, H., Kong, X., & Zhang, R. (2016, September). Fiber optic perimeter detection based on principal component analysis. In 2016 15th International Conference on Optical Communications and Networks (ICOCN) (pp. 1-3). IEEE.

[30] Wang, Z., Zhang, M., Wang, D., Song, C., Liu, M., Li, J., ... & Liu, Z. (2017). Failure prediction using machine learning and time series in optical network. Optics express, 25(16), 18553-18565.

[31] Zheng, C., Yu, S., & Gu, W. (2015, May). A SVM-based processor for free-space optical communication. In 2015 IEEE 5th International Conference on Electronics Information and Emergency Communication (pp. 30-33). IEEE.

[32] Wang, D., Zhang, M., Li, Z., Song, C., Fu, M., Li, J., & Chen, X. (2017). System impairment compensation in coherent optical communications by using a bio-inspired detector based on artificial neural network and genetic algorithm. Optics Communications, 399, 1-12.

[33] Villarrubia, G., De Paz, J. F., Chamoso, P., & De la Prieta, F. (2018). Artificial neural networks used in optimization problems. Neurocomputing, 272, 10-16.

[34] Wang, D., Zhang, M., Li, Z., Song, C., Fu, M., Li, J., & Chen, X. (2017). System impairment compensation in coherent optical communications by using a bio-inspired detector based on artificial neural network and genetic algorithm. *Optics Communications*, *399*, 1-12.

[35] Mizobuchi, T., Mizote, K., Kishikawa, H., Goto, N., & Yanagiya, S. I. (2012, July). QPSK label processing using complex-valued neural network learned with back propagation of teacher signals. In 2012 17th Opto-Electronics and Communications Conference (pp. 309-310). IEEE.

[36] Argyris, A., Bueno, J., & Fischer, I. (2018). Photonic machine learning implementation for signal recovery in optical communications. Scientific reports, 8(1), 8487.

[37] Lyubomirsky, I. (2015). Machine learning equalization techniques for high speed PAM4 fiber optic communication systems. CS229 Final Project Report, Stanford University.

[38] Ahmad, S. T., & Kumar, K. P. (2016). Radial basis function neural network nonlinear equalizer for 16-QAM coherent optical OFDM. IEEE Photonics Technology Letters, 28(22), 2507-2510.

[39] Giacoumidis, E., Le, S. T., Ghanbarisabagh, M., McCarthy, M., Aldaya, I., Mhatli, S., ... & Eggleton, B. J. (2015). Fiber nonlinearity-induced penalty reduction in CO-OFDM by ANN-based nonlinear equalization. Optics letters, 40(21), 5113-5116.

[40] de Sousa, T. F., & Fernandes, M. A. (2013, August). Multilayer perceptron equalizer for optical communication systems. In 2013 SBMO/IEEE MTT-S International Microwave & Optoelectronics Conference (IMOC) (pp. 1-5). IEEE.

[41] Liu, S., Wang, X., Zhang, W., Shen, G., & Tian, H. (2017). An Adaptive Activated ANN Equalizer Applied in Millimeter-Wave RoF Transmission System. IEEE Photonics Technology Letters, 29(22), 1935-1938.

[42] Liu, S., Alfadhli, Y. M., Shen, S., Xu, M., Tian, H., & Chang, G. K. (2018). A Novel ANN Equalizer to Mitigate Nonlinear Interference in Analog-RoF Mobile Fronthaul. IEEE Photonics Technology Letters, 30(19), 1675-1678.

[43] Jarajreh, M. A. K., Rajbhandari, S., Giacoumidis, E., Doran, N. J., & Ghassemlooy, Z. (2014, July). Fibre impairment compensation using artificial neural network equalizer for high-capacity coherent optical OFDM signals. In 2014 9th International Symposium on Communication Systems, Networks & Digital Sign (CSNDSP) (pp. 1112-1117). IEEE.

[44] Jarajreh, M. A., Giacoumidis, E., Aldaya, I., Le, S. T., Tsokanos, A., Ghassemlooy, Z., & Doran, N. J. (2015). Artificial neural network nonlinear equalizer for coherent optical OFDM. IEEE Photonics Technology Letters, 27(4), 387-390.

[45] Caballero, F. V., Ives, D. J., Laperle, C., Charlton, D., Zhuge, Q., O'Sullivan, M., & Savory, S. J. (2018). Machine learning based linear and nonlinear noise estimation. Journal of Optical Communications and Networking, 10(10), D42-D51.

[46] Samadi, P., Amar, D., Lepers, C., Lourdiane, M., & Bergman, K. (2017, September). Quality of transmission prediction with machine learning for dynamic operation of optical WDM networks. In 2017 European Conference on Optical Communication (ECOC) (pp. 1-3). IEEE.

[47] Wu, X., Jargon, J. A., Skoog, R. A., Paraschis, L., & Willner, A. E. (2009). Applications of artificial neural networks in optical performance monitoring. Journal of Lightwave Technology, 27(16), 3580-3589.

[48] Lai, J. S., Yang, A. Y., Zuo, L., & Sun, Y. N. (2011, November). Optical performance monitoring in 40-Gbps optical duobinary system using artificial neural networks trained with reconstructed eye diagram parameters. In Asia Communications and Photonics Conference and Exhibition (p. 83100M). Optical Society of America.

[49] Lai, J., Yang, A., & Sun, Y. (2011, August). Multiple-impairment monitoring for 40-Gbps RZ-OOK using artificial neural networks trained with reconstructed eye diagram parameters. In 2011 International Quantum Electronics Conference (IQEC) and Conference on Lasers and Electro-Optics (CLEO) Pacific Rim incorporating the Australasian Conference on Optics, Lasers and Spectroscopy and the Australian Conference on Optical Fibre Technology (pp. 563-565). IEEE.

[50] Wu, X., Jargon, J. A., Wang, C. M., & Willner, A. E. (2010, March). Experimental comparison of performance monitoring using neural networks trained with parameters derived from delay-tap plots and eye diagrams. In National Fiber Optic Engineers Conference (p. JThA17). Optical Society of America.

[51] Khan, F. N., Shen, T. S. R., Zhou, Y., Lau, A. P. T., & Lu, C. (2012). Optical performance monitoring using artificial neural networks trained with empirical moments of asynchronously sampled signal amplitudes. IEEE Photonics Technology Letters, 24(12), 982-984.

[52] Shen, T. S. R., Meng, K., Lau, A. P. T., & Dong, Z. Y. (2010). Optical performance monitoring using artificial neural network trained with asynchronous amplitude histograms. IEEE Photonics Technology Letters, 22(22), 1665-1667.

[53] Wu, X., Jargon, J. A., Paraschis, L., & Willner, A. E. (2011). ANN-based optical performance monitoring of QPSK signals using parameters derived from balanced-detected asynchronous diagrams. IEEE Photonics Technology Letters, 23(4), 248-250.

[54] Eriksson, T. A., Bülow, H., & Leven, A. (2017). Applying neural networks in optical communication systems: possible pitfalls. IEEE Photonics Technology Letters, 29(23), 2091-2094.

[55] Khan, F. N., Zhou, Y., Lau, A. P. T., & Lu, C. (2012). Modulation format identification in heterogeneous fiber-optic networks using artificial neural networks. Optics express, 20(11), 12422-12431.

[56] Wang, D., Zhang, M., Zhang, Z., Li, J., Gao, H., Zhang, F., & Chen, X. (2019). Machine Learning-Based Multifunctional Optical Spectrum Analysis Technique. IEEE Access, 7, 19726-19737.

[57] Barboza, E. D. A., Bastos-Filho, C. J., Martins-Filho, J. F., de Moura, U. C., & de Oliveira, J. R. (2013, August). Self-adaptive erbium-doped fiber amplifiers using machine learning. In 2013 SBMO/IEEE MTT-S International Microwave & Optoelectronics Conference (IMOC) (pp. 1-5). IEEE.

[58] Rafique, D., Szyrkowiec, T., Grießer, H., Autenrieth, A., & Elbers, J. P. (2018). Cognitive assurance architecture for optical network fault management. Journal of Lightwave Technology, 36(7), 1443-1450.

[59] Zhuge, Q., Zeng, X., Lun, H., Cai, M., Liu, X., Yi, L., & Hu, W. (2019). Application of Machine Learning in Fiber Nonlinearity Modeling and Monitoring for Elastic Optical Networks. Journal of Lightwave Technology.

[60] Proietti, R., Chen, X., Zhang, K., Liu, G., Shamsabardeh, M., Castro, A., ... & Yoo, S. B. (2019). Experimental demonstration of machine-learning-aided QoT estimation in multi-domain elastic optical networks with alien wavelengths. IEEE/OSA Journal of Optical Communications and Networking, 11(1), A1-A10.

[61] Yan, S., Khan, F. N., Mavromatis, A., Gkounis, D., Fan, Q., Ntavou, F., ... & Lu, C. (2017, September). Field trial of machine-learning-assisted and SDN-based optical network planning with network-scale monitoring database. In 2017 European Conference on Optical Communication (ECOC) (pp. 1-3). IEEE.

[62] Zhou, X., Zhuge, Q., Qiu, M., Xiang, M., Zhang, F., Wu, B., ... & Plant, D. V. (2018). Bandwidth variable transceivers with artificial neural network-aided provisioning and capacity improvement capabilities in meshed optical networks with cascaded ROADM filtering. *Optics Communications*, *409*, 23-33.

[63] Bensalem, M., Singh, S. K., & Jukan, A. (2019). Machine Learning Techniques to Detecting and Preventing Jamming Attacks in Optical Networks. *arXiv preprint arXiv:1902.07537*.

[64] Aibin, M. (2018). Traffic prediction based on machine learning for elastic optical networks. Optical Switching and Networking, 30, 33-39.

[65] Ruan, L., & Wong, E. (2018, May). Machine intelligence in allocating bandwidth to achieve low-latency performance. In 2018 International Conference on Optical Network Design and Modeling (ONDM) (pp. 226-229). IEEE.

[66] Li, Z., & Zhao, X. (2017). BP artificial neural network based wave front correction for sensor-less free space optics communication. Optics Communications, 385, 219-228.

[67] Rajbhandari, S., Haigh, P. A., Ghassemlooy, Z., & Popoola, W. (2013). Wavelet-neural network VLC receiver in the presence of artificial light interference. IEEE photonics technology letters, 25(15), 1424-1427.

[68] Haigh, P. A., Ghassemlooy, Z., Rajbhandari, S., Papakonstantinou, I., & Popoola, W. (2014). Visible light communications: 170 Mb/s using an artificial neural network equalizer in a low bandwidth white light configuration. Journal of lightwave technology, 32(9), 1807-1813.

[69] Haigh, P. A., Ghassemlooy, Z., Papakonstantinou, I., & Rajbhandari, S. (2013, July). Online artificial neural network equalization for a visible light communications system with an organic light emitting diode based transmitter. In Proceedings of the 2013 18th European Conference on Network and Optical Communications & 2013 8th Conference on Optical Cabling and Infrastructure (NOC-OC&I) (pp. 153-158). IEEE.

[70] Haigh, P. A., Ghassemlooy, Z., Papakonstantinou, I., Tedde, F., Tedde, S. F., Hayden, O., & Rajbhandari, S. (2013, June). A MIMO-ANN system for increasing data rates in organic visible light communications systems. In 2013 IEEE International Conference on Communications (ICC) (pp. 5322-5327). IEEE.

[71] Gao, D., Guo, Q. (2019). Extreme Learning Machine-Based Receiver for MIMO LED Communications, arXiv:1903.01551

[72] Leung, H. C., Leung, C. S., Wong, E. W., & Li, S. (2017). Extreme learning machine for estimating blocking probability of bufferless OBS/OPS networks. IEEE/OSA Journal of Optical Communications and Networking, 9(8), 682-692.

[73] Xu, T., Xu, T., & Darwazeh, I. (2018, August). Deep learning for interference cancellation in non-orthogonal signal based optical communication systems. In 2018 Progress in Electromagnetics Research Symposium (PIERS-Toyama) (pp. 241-248). IEEE.

[74] Amirabadi, M. A. (2019). Novel Suboptimal approaches for Hyperparameter Tuning of Deep Neural Network [under the shelf of Optical Communication]. arXiv preprint arXiv:1907.00036.

[75] Giacoumidis, E., Wei, J., Aldaya, I., & Barry, L. P. (2018). Exceeding the nonlinear Shannon-limit in coherent optical communications using 3D adaptive machine learning. arXiv preprint arXiv:1802.09120.

[76] Häger, C., & Pfister, H. D. (2018, March). Nonlinear interference mitigation via deep neural networks. In 2018 Optical Fiber Communications Conference and Exposition (OFC) (pp. 1-3). IEEE.

[77] Poudel, B., Oshima, J., Kobayashi, H., & Iwashita, K. (2019). MIMO detection using a deep learning neural network in a mode division multiplexing optical transmission system. Optics Communications, 440, 41-48.

[78] Xiong, Y., Yang, Y., Ye, Y., & Rouskas, G. N. (2019). A machine learning approach to mitigating fragmentation and crosstalk in space division multiplexing elastic optical networks. Optical Fiber Technology, 50, 99-107.

[79] Khan, F. N., Zhong, K., Al-Arashi, W. H., Yu, C., Lu, C., & Lau, A. P. T. (2016). Modulation format identification in coherent receivers using deep machine learning. IEEE Photonics Technology Letters, 28(17), 1886-1889.

[80] Khan, F. N., Zhong, K., Zhou, X., Al-Arashi, W. H., Yu, C., Lu, C., & Lau, A. P. T. (2017). Joint OSNR monitoring and modulation format identification in digital coherent receivers using deep neural networks. Optics express, 25(15), 17767-17776.

[81] Karanov, B., Chagnon, M., Thouin, F., Eriksson, T. A., Bülow, H., Lavery, D., ... & Schmalen, L. (2018). End-to-end deep learning of optical fiber communications. Journal of Lightwave Technology, 36(20), 4843-4855.

[82] Karanov, B., Lavery, D., Bayvel, P., & Schmalen, L. (2019). End-to-End Optimized Transmission over Dispersive Intensity-Modulated Channels Using Bidirectional Recurrent Neural Networks. arXiv preprint arXiv:1901.08570.

[83] O'Shea, T., & Hoydis, J. (2017). An introduction to deep learning for the physical layer. IEEE Transactions on Cognitive Communications and Networking, 3(4), 563-575.

[84] Jones, R. T., Eriksson, T. A., Yankov, M. P., & Zibar, D. (2018, September). Deep learning of geometric constellation shaping including fiber nonlinearities. In 2018 European Conference on Optical Communication (ECOC) (pp. 1-3). IEEE.

[85] Li, S., Häger, C., Garcia, N., & Wymeersch, H. (2018, September). Achievable information rates for nonlinear fiber communication via end-to-end autoencoder learning. In 2018 European Conference on Optical Communication (ECOC) (pp. 1-3). IEEE.

[86] Jones, R. T., Eriksson, T. A., Yankov, M. P., Puttnam, B. J., Rademacher, G., Luis, R. S., & Zibar, D. (2018). Geometric constellation shaping for fiber optic communication systems via end-to-end learning. arXiv preprint arXiv:1810.00774.

[87] Wang, C., Fu, S., Xiao, Z., Tang, M., & Liu, D. (2019). Long short-term memory neural network (LSTM-NN) enabled accurate optical signal-to-noise ratio (OSNR) monitoring. Journal of Lightwave Technology.

[88] Zibar, D., Ferrari, A., Curri, V., & Carena, A. (2019, March). Machine learning-based Raman amplifier design. In Optical Fiber Communication Conference (pp. M1J-1). Optical Society of America.

[89] Zhong, Z., Hua, N., Yuan, Z., Li, Y., & Zheng, X. (2019, March). Routing without Routing Algorithms: an AI-Based Routing Paradigm for Multi-Domain Optical Networks. In Optical Fiber Communication Conference (pp. Th2A-24). Optical Society of America.

[90] Tian, Q., Li, Z., Hu, K., Zhu, L., Pan, X., Zhang, Q., ... & Xin, X. (2018). Turbo-coded 16-ary OAM shift keying FSO communication system combining the CNN-based adaptive demodulator. Optics express, 26(21), 27849-27864.

[91] Wang, D., Zhang, M., Li, J., Li, Z., Li, J., Song, C., & Chen, X. (2017). Intelligent constellation diagram analyzer using convolutional neural network-based deep learning. Optics express, 25(15), 17150-17166.

[92] Zhang, J., Chen, W., Gao, M., Ma, Y., Zhao, Y., & Shen, G. (2018). Intelligent adaptive coherent optical receiver based on convolutional neural network and clustering algorithm. Optics express, 26(14), 18684-18698.

[93] Tanimura, T., Hoshida, T., Kato, T., Watanabe, S., & Morikawa, H. (2019). Convolutional neural network-based optical performance monitoring for optical transport networks. Journal of Optical Communications and Networking, 11(1), A52-A59.

[94] Panayiotou, T., Savva, G., Shariati, B., Tomkos, I., & Ellinas, G. (2019, March). Machine Learning for QoT Estimation of Unseen Optical Network States. In Optical Fiber Communication Conference (pp. Tu2E-2). Optical Society of America.

[95] Cuevas, A. R., Fontana, M., Rodriguez-Cobo, L., Lomer, M., & López-Higuera, J. M. (2018). Machine Learning for Turning Optical Fiber Specklegram Sensor into a Spatially-Resolved Sensing System. Proof of Concept. Journal of Lightwave Technology, 36(17), 3733-3738.

[96] Li, J., Zhang, M., Wang, D., Wu, S., & Zhan, Y. (2018). Joint atmospheric turbulence detection and adaptive demodulation technique using the CNN for the OAM-FSO communication. Optics express, 26(8), 10494-10508.

[97] Wang, Z., Dedo, M. I., Guo, K., Zhou, K., Shen, F., Sun, Y., ... & Guo, Z. (2019). Efficient recognition of the propagated orbital angular momentum modes in turbulences with the convolutional neural network. IEEE Photonics Journal.

[98] Ma, S., Dai, J., Lu, S., Li, H., Zhang, H., Du, C., & Li, S. (2019). Signal Demodulation With Machine Learning Methods for Physical Layer Visible Light Communications: Prototype Platform, Open Dataset, and Algorithms. IEEE Access, 7, 30588-30598.

[99] Wang, K., Yu, X., Xiong, Q., Zhu, Q., Lu, W., Huang, Y., & Zhao, L. (2019). Learning to improve WLAN indoor positioning accuracy based on DBSCAN-KRF algorithm from RSS fingerprint data. *IEEE Access*.

[100] Jiang, L., Yan, L., Yi, A., Pan, Y., Pan, W., & Luo, B. (2018). K-Nearest Neighbor Detector for Enhancing Performance of Optical Phase Conjugation System in the Presence of Nonlinear Phase Noise. IEEE Photonics Journal, 10(2), 1-8.

[101] Wang, D., Zhang, M., Fu, M., Cai, Z., Li, Z., Han, H., ... & Luo, B. (2016). Nonlinearity mitigation using a machine learning detector based on $k$-nearest neighbors. IEEE Photonics Technology Letters, 28(19), 2102-2105.

[102] Zhang, J., Gao, M., Chen, W., & Shen, G. (2018). Non-data-aided k-nearest neighbors technique for optical fiber nonlinearity mitigation. Journal of Lightwave Technology, 36(17), 3564-3572.

[103] Barletta, L., Giusti, A., Rottondi, C., & Tornatore, M. (2017, March). QoT estimation for unestablished lighpaths using machine learning. In Optical Fiber Communication Conference(pp. Th1J-1). Optical Society of America.

[104] Rottondi, C., Barletta, L., Giusti, A., & Tornatore, M. (2018). Machine-learning method for quality of transmission prediction of unestablished lightpaths. IEEE/OSA Journal of Optical Communications and Networking, 10(2), A286-A297.

[105] Tóth, J., Ovseník, Ľ., Turán, J., Michaeli, L., & Márton, M. (2018). Classification prediction analysis of RSSI parameter in hard switching process for FSO/RF systems. Measurement, 116, 602-610.

[106] Sorokina, M., Sygletos, S., & Turitsyn, S. (2016). Sparse identification for nonlinear optical communication systems: SINO method. Optics express, 24(26), 30433-30443.

[107] Sorokina, M., Sygletos, S., & Turitsyn, S. (2017, July). Sparse Identification for Nonlinear Optical communication systems. In 2017 19th International Conference on Transparent Optical Networks (ICTON) (pp. 1-4). IEEE.

[108] Seve, E., Pesic, J., Delezoide, C., Bigo, S., & Pointurier, Y. (2018). Learning process for reducing uncertainties on network parameters and design margins. Journal of Optical Communications and Networking, 10(2), A298-A306.

[109] Huang, Y., Gutterman, C. L., Samadi, P., Cho, P. B., Samoud, W., Ware, C., ... & Bergman, K. (2017). Dynamic mitigation of EDFA power excursions with machine learning. Optics express, 25(3), 2245-2258.

[110] Huang, Y., Cho, P. B., Samadi, P., & Bergman, K. (2017, March). Dynamic power pre-adjustments with machine learning that mitigate EDFA excursions during defragmentation. In 2017 Optical Fiber Communications Conference and Exhibition (OFC)(pp. 1-3). IEEE.

[111] Huang, Y., Cho, P. B., Samadi, P., & Bergman, K. (2018). Power excursion mitigation for flexgrid defragmentation with machine learning. Journal of Optical Communications and Networking, 10(1), A69-A76.

[112] Tran, H. Q., & Ha, C. (2019). Improved Visible Light-Based Indoor Positioning System Using Machine Learning Classification and Regression. Applied Sciences, 9(6), 1048.

[113] Jones, R. T. (2019). Machine Learning Methods in Coherent Optical Communication Systems.

[114] Mai, X., Liu, J., Wu, X., Zhang, Q., Guo, C., Yang, Y., & Li, Z. (2017). Stokes space modulation format classification based on non-iterative clustering algorithm for coherent optical receivers. Optics express, 25(3), 2038-2050.

[115] Cheng, L., Xi, L., Zhao, D., Tang, X., Zhang, W., & Zhang, X. (2015). Improved modulation format identification based on Stokes parameters using combination of fuzzy c-means and hierarchical clustering in coherent optical communication system. Chinese Optics Letters, 13(10), 100604.

[116] Georgoulakis, K., Matrakidis, C., Glentis, G. O., & Stavdas, A. (2010, June). Fractionally spaced clustering based equalizer for optical channels. In Signal Processing in Photonic Communications (p. SPWC3). Optical Society of America.

[117] Torres, J. J. G., Varughese, S., Thomas, V. A., Chiuchiarelli, A., Ralph, S. E., Soto, A. M. C., & González, N. G. (2017). Mitigation of time-varying distortions in Nyquist-WDM systems using machine learning. Optical Fiber Technology, 38, 130-135.

[118] Giacoumidis, E., Lin, Y., & Barry, L. P. (2019). DBSCAN for nonlinear equalization in high-capacity multi-carrier optical communications. arXiv preprint arXiv:1902.01198.

[119] Amari, A., Lin, X., Dobre, O. A., Venkatesan, R., & Alvarado, A. (2019). A Machine Learning-Based Detection Technique for Optical Fiber Nonlinearity Mitigation. IEEE Photonics Technology Letters.

[120] Zhang, J., Chen, W., Gao, M., & Shen, G. (2017). K-means-clustering-based fiber nonlinearity equalization techniques for 64-QAM coherent optical communication system. Optics express, 25(22), 27570-27580.

[121] Zhang, L., Pang, X., Ozolins, O., Udalcovs, A., Popov, S., Xiao, S., ... & Chen, J. (2018). Spectrally efficient digitized radio-over-fiber system with k-means clustering-based multidimensional quantization. Optics letters, 43(7), 1546-1549.

[122] Gonzalez, N. G., Zibar, D., Yu, X., & Monroy, I. T. (2010, March). Optical phase-modulated radio-over-fiber links with k-means algorithm for digital demodulation of 8PSK subcarrier multiplexed signals. In Optical Fiber Communication Conference (p. OML3). Optical Society of America.

[123] Gonzalez, N. G., Zibar, D., Caballero, A., & Monroy, I. T. (2010). Experimental 2.5-Gb/s QPSK WDM Phase-Modulated Radio-Over-Fiber Link With Digital Demodulation by a $K$-Means Algorithm. IEEE Photonics Technology Letters, 22(5), 335-337.

[124] Chen, H., Li, Y., & Shen, G. (2015, November). Planning for passive optical network deployment with K-means clustering-based approach. In Asia Communications and Photonics Conference (pp. AM2E-2). Optical Society of America.

[125] Wu, X., & Chi, N. (2019). The phase estimation of geometric shaping 8-QAM modulations based on K-means clustering in underwater visible light communication. Optics Communications.

[126] Lu, X., Zhou, Y., Qiao, L., Yu, W., Liang, S., Zhao, M., ... & Chi, N. (2019). Amplitude jitter compensation of PAM-8 VLC system employing time-amplitude two-Dimensional re-estimation base on density clustering of machine learning. Physica Scripta.

[127] Zibar, D., de Carvalho, L. H. H., Piels, M., Doberstein, A., Diniz, J., Nebendahl, B., ... & de Oliveira, J. C. R. (2015). Application of machine learning techniques for amplitude and phase noise characterization. Journal of Lightwave Technology, 33(7), 1333-1343.

[128] Zibar, D., Piels, M., Jones, R., & Schäeffer, C. G. (2016). Machine learning techniques in optical communication. Journal of Lightwave Technology, 34(6), 1442-1452.

[129] Tan, M. C., Khan, F. N., Al-Arashi, W. H., Zhou, Y., & Lau, A. P. T. (2014). Simultaneous optical performance monitoring and modulation format/bit-rate identification using principal component analysis. Journal of Optical Communications and Networking, 6(5), 441-448.

[130] Yang, S., Zhang, X., Xi, L., & Lu, C. (2011). Optical network traffic detectional algorithm based on principal component analysis.

[131] Alaíz, C. M., Fanuel, M., & Suykens, J. A. (2018). Convex formulation for kernel PCA and its use in semisupervised learning. IEEE transactions on neural networks and learning systems, 29(8), 3863-3869.

[132] Lu, H., Cui, S., Ke, C., & Liu, D. (2017). Automatic reference optical spectrum retrieval method for ultra-high resolution optical spectrum distortion analysis utilizing integrated machine learning techniques. Optics Express, 25(26), 32491-32503.

[133] Razo-Zapata, I. S., Castañón, G., & Mex-Perera, C. (2014). Self-healing in transparent optical packet switching mesh networks: A reinforcement learning perspective. Computer Networks, 60, 129-146.

[134] Belbekkouche, A., & Hafid, A. (2007, June). An adaptive reinforcement learning-based approach to reduce blocking probability in bufferless OBS networks. In 2007 IEEE International Conference on Communications (pp. 2377-2382). IEEE.

[135] Belbekkouche, A., Hafid, A., & Gendreau, M. (2008, November). A reinforcement learning-based deflection routing scheme for buffer-less OBS networks. In IEEE GLOBECOM 2008-2008 IEEE Global Telecommunications Conference (pp. 1-6). IEEE.

[136] Belbekkouche, A., Hafid, A., & Gendreau, M. (2009, September). Adaptive Routing and Contention Resolution approaches for OBS networks with QoS differentiation. In 2009 Sixth International Conference on Broadband Communications, Networks, and Systems (pp. 1-8). IEEE.

[137] Kiran, Y. V., Venkatesh, T., & Murthy, C. S. R. (2009, June). A multi-agent reinforcement learning approach to path selection in optical burst switching networks. In 2009 IEEE International Conference on Communications (pp. 1-5). IEEE.

[138] Pointurier, Y., & Heidari, F. (2007, September). Reinforcement learning based routing in all-optical networks. In 2007 Fourth International Conference on Broadband Communications, Networks and Systems (BROADNETS'07) (pp. 919-921). IEEE.

[139] Razo-Zapata, I. S., Castañon, G., & Mex-Perera, C. (2013, June). Lightpath requests processing in flexible packet switching optical networks using reinforcement learning. In 2013 15th International Conference on Transparent Optical Networks (ICTON) (pp. 1-4). IEEE.

[140] Garcia, P., Zsigri, A., & Guitton, A. (2003, June). A multicast reinforcement learning algorithm for WDM optical networks. In Proceedings of the 7th International Conference on Telecommunications, 2003. ConTEL 2003. (Vol. 2, pp. 419-426). IEEE.

[141] Heidari, F., Mannor, S., & Mason, L. G. (2007, May). Reinforcement learning-based load shared sequential routing. In *International Conference on Research in Networking* (pp. 832-843). Springer, Berlin, Heidelberg.

[142] Natalino, C., Raza, M. R., Batista, P., Santos, M., Wosinska, L., & Monti, P. (2018). Machine-learning-based routing of QoS-constrained connectivity services in optical transport networks. *Proc. of OSA Advanced Photonics], NeW3F*, *5*.

[143] Seve, E., Pesic, J., Delezoide, C., Bigo, S., & Pointurier, Y. (2018). Learning process for reducing uncertainties on network parameters and design margins. *Journal of Optical Communications and Networking*, *10*(2), A298-A306.

[144] Bkassiny, M., Jayaweera, S. K., & Avery, K. A. (2011). Distributed Reinforcement Learning based MAC protocols for autonomous cognitive secondary users. NEW MEXICO UNIV ALBUQUERQUE DEPT OF ELECTRICAL ENGINEERING AND COMPUTER SCIENCE.

[145] Du, Z., Wang, C., Sun, Y., & Wu, G. (2018). Context-Aware Indoor VLC/RF Heterogeneous Network Selection: Reinforcement Learning with Knowledge Transfer. IEEE Access, 6, 33275-33284.

[146] Kiran, Y. V., Venkatesh, T., & Murthy, C. S. R. (2007). A reinforcement learning framework for path selection and wavelength selection in optical burst switched networks. IEEE Journal on Selected Areas in Communications, 25(9), 18-26.

[147] Haeri, S., Thong, W. W. K., Chen, G., & Trajković, L. (2013, August). A reinforcement learning-based algorithm for deflection routing in optical burst-switched networks. In 2013 IEEE 14th International Conference on Information Reuse & Integration (IRI) (pp. 474-481). IEEE.

[148] Kiran, Y. V., Venkatesh, T., & Murthy, C. S. R. (2006, October). Reinforcement learning based path selection and wavelength selection in optical burst switched networks. In 2006 3rd International Conference on Broadband Communications, Networks and Systems (pp. 1-8). IEEE.

[149] Liu, Y., Yang, Y., Han, P., Shao, Z., & Li, C. (2019). Virtual Network Embedding in Fiber-Wireless Access Networks for Resource-Efficient IoT Service Provisioning. *IEEE Access*, *7*, 65506-65517.

[150] Luo, X., Shi, C., Wang, L., Chen, X., Li, Y., & Yang, T. (2019). Leveraging double-agent-based deep reinforcement learning to global optimization of elastic optical networks with enhanced survivability. Optics Express, 27(6), 7896-7911.

[151] Lu, W., Fang, H., & Zhu, Z. (2018, May). AI-assisted resource advertising and pricing to realize distributed tenant-driven virtual network slicing in inter-DC optical networks. In 2018 International Conference on Optical Network Design and Modeling (ONDM)(pp. 130-135). IEEE.

[152] Mata, J., De Miguel, I., Duran, R. J., Merayo, N., Singh, S. K., Jukan, A., & Chamania, M. (2018). Artificial intelligence (AI) methods in optical networks: A comprehensive survey. Optical Switching and Networking, 28, 43-57.

[153] Musumeci, F., Rottondi, C., Nag, A., Macaluso, I., Zibar, D., Ruffini, M., & Tornatore, M. (2018). An overview on application of machine learning techniques in optical networks. IEEE Communications Surveys & Tutorials.

[154] Rafique, D., & Velasco, L. (2018). Machine learning for network automation: overview, architecture, and applications [Invited Tutorial]. Journal of Optical Communications and Networking, 10(10), D126-D143.

[155] Giacoumidis, E., Lin, Y., Wei, J., Aldaya, I., Tsokanos, A., & Barry, L. (2019). Harnessing machine learning for fiber-induced nonlinearity mitigation in long-haul coherent optical OFDM. Future Internet, 11(1), 2.

[156] Khan, F. N., Fan, Q., Lu, C., & Lau, A. P. T. (2019). An Optical Communication's Perspective on Machine Learning and Its Applications. Journal of Lightwave Technology, 37(2), 493-516.